# Optimization-Based Discovery of A Non-Attracting Flow State in An Oscillating-Cylinder Wake

Daiwei Dong[1], Wenbo Cao[1], Wei Suo[1], Jiaqing Kou[1], Weiwei Zhang[1,2,3,*]

1.School of Aeronautic, Northwestern Polytechnical University, Xi'an 710072, China;

2.International Joint Institute of Artificial Intelligence on Fluid Mechanics,

Northwestern Polytechnical University, Xi'an, 710072, China;

3.National Key Laboratory of Aircraft Configuration Design, Xi'an 710072, China

## Abstract

Nonlinear dynamical systems governed by differential equations often admit multiple solutions, including states that satisfy the governing equations but are dynamically non-attracting. Such solutions have been extensively studied in low-dimensional systems, but remain far less explored in more complex systems, particularly in fluid flows governed by the Navier–Stokes equations.

A classical example is the unstable steady solution underlying the Kármán vortex street. This solution is time-independent and non-attracting: although it satisfies the governing equations, it is unstable and typically inaccessible to direct time-stepping simulations. This naturally raises the question of whether analogous time-dependent non-attracting solutions exist—solutions that satisfy the governing equations but remain inaccessible to conventional time-stepping methods.

In the present study, while employing physics-informed neural networks (PINNs) to solve the incompressible Navier–Stokes equations and investigate the flow past a forced oscillating cylinder at supercritical Reynolds numbers, we identify a class of flow solutions that are inaccessible through direct time-stepping simulations. The obtained solution exhibits a single dominant frequency and remains phase-locked with the cylinder oscillation frequency, despite the corresponding parameters lying outside the conventional lock-in regime.

To verify this solution, the obtained PINNs solution is used as the initial guess for an optimization based on the optimizing a discrete loss (ODIL) framework. The results show that the solution can be consistently maintained during the optimization process while satisfying the governing equations and oscillatory boundary conditions. This indicates that the solution is self-consistent in the optimization sense, although it does not correspond to an attracting state of the original dynamical system.

To understand why such solutions can be obtained, we compare the numerical evolution mechanisms of conventional time-stepping methods and optimization-based solvers. The convergence of time-stepping methods is governed by the spectral properties of the Jacobian operator $A$ obtained from the linearization of the governing equations. In contrast, optimization-based methods minimize the quadratic norm of the residual of the governing equations, and the associated gradient evolution is governed by the normal operator $A^T A$. Because this operator does not inherit the unstable spectral properties of the original dynamical operator, flow states that satisfy the governing equations but are dynamically non-attracting can be identified and maintained as minima of the optimization problem.

These results indicate that, for the flow past a forced oscillating cylinder, non-attracting periodic solutions that satisfy the governing equations exist in addition to the attracting states obtained by conventional time-stepping simulations. Optimization-based solvers can therefore reveal such flow states that are difficult to obtain through direct time integration, providing a new perspective for understanding complex wake dynamics.

# 1 Introduction

The flow past a circular cylinder has long been regarded as a canonical paradigm for studying bluff-body wakes. Its dynamics are primarily governed by the Reynolds number $Re = U_\infty d / \nu$, where $U_\infty$ is the freestream velocity, $d$ is the cylinder diameter, and $\nu$ is the kinematic viscosity. At low Reynolds numbers, when $Re$ is below a critical value $Re_c \approx 46$, the flow remains a two-dimensional steady state. As $Re$ approaches $Re_c$, the steady solution undergoes a Hopf bifurcation and loses stability, giving rise to a time-periodic wake with a well-defined dominant frequency, known as the classical Bénard–von Kármán vortex street [1-3]. With a further increase in Reynolds number, when $Re_c \approx 188$, the two-dimensional periodic flow becomes three-dimensionally unstable and transitions to a fully three-dimensional unsteady state[4]. The vortex street arising from the Hopf bifurcation is one of the most fundamental unsteady flows in fluid mechanics and serves as a benchmark model for understanding nonlinear dynamics and bifurcation phenomena in bluff-body wakes.

It is worth noting that, although the steady solution becomes inaccessible to direct time-stepping beyond the critical Reynolds number, it remains an exact solution of the governing equations and plays a central role in understanding the formation and stability of the vortex street. Classical studies have shown that the two-dimensional steady wake undergoes a Hopf bifurcation near the critical Reynolds number, leading to periodic vortex shedding, while linearization about this steady base flow reveals the dominant instability modes and the associated frequency selection mechanism [5]. The work of Noack and Eckelmann [6] systematically examined the relationship between steady and periodic solutions in cylinder wakes. By performing global linear stability analysis using the unstable two-dimensional steady solution as the base flow, they demonstrated that a supercritical Hopf bifurcation of this base flow determines the onset of periodic vortex shedding at the critical Reynolds number. Furthermore, they accurately predicted the critical conditions for the subsequent three-dimensional instability at higher Reynolds numbers, highlighting the key role of the unstable steady solution in the nonlinear evolution of the wake. In fluid–structure interaction problems, several studies have similarly used unstable steady solutions as base flows for global linear stability analysis, successfully explaining the origin of complex oscillatory phenomena such as lock-in and galloping [7] [8] [9]. Likewise, in flow control studies, unstable steady solutions are widely adopted as base flows to analyze disturbance growth and flow sensitivity to control strategies, including wake control [10], vortex-induced vibration (VIV) [11] [12], and flutter control [13]. By linearizing the flow about such steady solutions, one can identify dominant modes, frequency selection mechanisms, and energy transfer pathways, thereby providing a theoretical basis for control design.

These studies collectively indicate that, although the unstable steady solution is not dynamically attracting and cannot be obtained through time integration, it serves as indispensable reference states for theoretical analysis, modal decomposition, and control design, and plays a fundamental role in understanding unsteady wake dynamics and stability boundaries.

To investigate such flow problems, partial differential equations provide the fundamental mathematical framework for describing fluid motion. Conventional numerical methods typically begin by discretizing the governing equations in space and time, for example using finite difference [14], finite volume [15], or finite element methods [16], thereby transforming the continuous problem into a system of algebraic equations. The resulting discrete system is then solved using iterative solvers such as Gauss–Seidel, Jacobi, successive over-relaxation (SOR), and the generalized minimal residual (GMRES) method [17] [18]. These approaches essentially rely on time marching or pseudo-time evolution, and their convergence and stability are directly constrained by the spectral properties of the resulting discrete dynamical system.

Building on this foundation, a class of methods has emerged in recent years that revisits the numerical solution of partial differential equations from an optimization perspective. These approaches obtain solutions by minimizing an objective function constructed from the residuals of the discretized governing equations. A representative example is the Optimizing a Discrete Loss (ODIL) method [19]. This approach still adopts standard grid-based discretization, but does not explicitly rely on time stepping. Instead, it treats the discretized governing equations as constraints and minimizes the $L^2$-norm of their residuals using gradient-based optimization algorithms. This strategy inherently preserves the numerical consistency and conservation properties of classical discrete schemes, allowing ODIL to maintain close ties to traditional numerical methods in terms of accuracy and physical fidelity. Simultaneously, its numerical evolution mechanism exhibits characteristics distinctly different from those of time-iteration methods.

Meanwhile, artificial neural networks (ANNs) and their deep variants have demonstrated strong approximation capabilities in scientific computing and have been widely applied to solve the partial differential equations. Unlike traditional grid-based numerical methods, these approaches parameterize the solution of the governing equations using deep neural networks, thereby transforming the solution process into an optimization problem over the network parameters. Training such neural-network-based PDE solvers is typically achieved by minimizing a loss function composed of the residual of the governing equations together with the boundary and initial conditions. Depending on how the governing equations are formulated, existing approaches can be broadly classified into methods based on the strong form, such as physics-informed neural networks (PINNs) [20] and the deep Galerkin method [21], and methods based on weak or variational forms, including the deep Ritz method [22], the deep Nitsche method [23], and the deep energy method [24].

This study focuses on physics-informed neural networks (PINNs). In fluid mechanics, the PINN

framework has been applied to a variety of representative problems. Sun et al. [25] solved parameterized Navier–Stokes equations using PINNs, systematically investigating the effects of physical parameters such as viscosity on flow solutions. Subsequently, Sun et al. [26] applied this approach to low-Reynolds-number flow over airfoils, demonstrating its capability in handling parameterized geometries and flow prediction. For inviscid flows, Cao et al. [27] [28] combined PINNs with mesh transformation techniques to solve parameterized Euler equations, incorporating both flow conditions and geometric configurations as input parameters. Building on this work, Cao et al. further proposed a time-stepping-oriented neural network (TSONN) framework [29], which was successfully applied to the lid-driven cavity flow at a Reynolds number of 5000, as well as two-dimensional laminar flow around airfoils and three-dimensional laminar flow around wings [30]. More recently, this approach has been extended to high-Reynolds-number wall-bounded turbulent flows governed by the Reynolds-averaged Navier–Stokes equations coupled with the Spalart–Allmaras turbulence model [31]. For flows involving moving boundaries, Zhu et al. [32] developed a PINN-based method for incompressible flows with time-dependent moving geometries. Their work covers a range of benchmark cases, from simple single-body oscillations to complex multi-body motions, and from purely physics-driven setups to scenarios incorporating partial data for reconstruction, demonstrating the applicability of the framework to flows with time-varying boundaries.

Regarding multiple-solution problems, existing studies have shown that PINNs can be viewed as a global optimization framework based on minimizing the residual of the governing equations, and therefore have the potential to explore multiple solution structures of nonlinear partial differential equations. The HomPINNs method [33] [34] embeds a homotopy parameter into the PINN training process, enabling continuous tracking of multiple solution branches of nonlinear elliptic equations. It has successfully identified multiple steady solutions in several benchmark cases, demonstrating the capability of the optimization-based PINN framework to discover multiple solutions without explicit continuation from initial guesses. Building on this, further studies indicate that PINNs can not only reconstruct known solution branches, but also identify non-attracting solution structures that are difficult to capture using conventional time-stepping or Newton-type methods. In particular, Wang et al. [35] employed high-accuracy PINNs to systematically investigate several nonlinear fluid models, and for the first time numerically identified and analyzed families of unstable singular solutions, revealing their consistency with the governing equations and their non-attracting nature in the dynamical system. This highlights the unique capability of PINNs in exploring complex solution spaces of PDEs. Meanwhile, the ability of PINNs to capture multiple solutions has also been demonstrated in classical fluid mechanics problems. Zou et al. [36] successfully obtained multiple distinct flow solutions satisfying the same governing equations and boundary conditions in canonical cases such as lid-driven cavity flow, using strategies of random initialization and deep ensembles. These results further indicate that optimization based on residual minimization can overcome the accessibility limitations of traditional time-stepping methods, providing a new numerical approach for characterizing multiple-solution structures in complex flow problems.

In the flow past a forced oscillating cylinder, the introduction of external forcing leads to a fundamentally unsteady system, in which the flow response arises from the interaction between the natural vortex shedding and the imposed oscillation. To interpret the lock-in mechanism from a dynamical systems perspective, it is common to employ low-dimensional models based on the normal form near a Hopf bifurcation, such as the Stuart–Landau equation [3, 37, 38]. Under external periodic forcing, this model can be further reduced to a phase equation (e.g., the Adler equation), which describes the evolution of the phase difference between the flow response and the external forcing. Within this framework, phase locking refers to a state in which this phase difference remains constant in time, corresponding to a stable equilibrium in the phase dynamics and resulting in a single-frequency response synchronized with the forcing. Outside the lock-in region, no stable equilibrium exists, and the phase difference continuously drifts, leading to a response involving multiple frequencies. However, this does not necessarily imply the nonexistence of the corresponding single-frequency solution. Instead, such a solution, if it exists, becomes dynamically non-attracting and therefore cannot be obtained through conventional time-stepping simulations. This suggests that, in forced oscillatory flows, there may also exist flow states that satisfy the governing equations but remain hidden from standard numerical approaches.

In this work, while using PINNs to solve the incompressible Navier–Stokes equations and study the flow past a forced oscillating cylinder at supercritical Reynolds numbers, we identify a class of flow solutions that are not accessible through direct time-stepping methods. The solution exhibits a single dominant frequency and remains phase-locked with the cylinder oscillation, even though the corresponding parameters lie outside the conventional lock-in regime. To verify the solution, the PINN result is used as the initial guess in ODIL framework. The solution is maintained throughout the optimization process, while satisfying the governing equations and oscillatory boundary conditions, indicating that it is self-consistent in the optimization sense, although it is not an attracting state of the original dynamical system. To understand the numerical mechanism behind the existence of such solutions, we compare the evolution behavior of conventional time-stepping methods and optimization-based solvers. The analysis shows that flow states that strictly satisfy the governing equations but are dynamically non-attracting can still be identified and maintained as minima of the optimization problem.

The remainder of this paper is organized as follows. Section 2 reviews the incompressible Navier–

Stokes equations and the formulation of the forced oscillating cylinder problem, and introduces the PINNs and ODIL methods. In Section 3, the flow solution obtained by PINNs is used as the initial guess in the ODIL framework for further computation, and the results are analyzed and discussed. Section 4 provides a systematic comparison between conventional time-stepping methods and optimization-based solvers from the perspective of numerical evolution, in order to explain the observed phenomena. Finally, Section 5 summarizes the main findings of this work. In the Appendix, a simple Hopf bifurcation problem is solved using PINNs to further illustrate this difference.

## 2 Method
### 2.1 Governing Equations and Discrete Solution Methods

In a Cartesian coordinate system, the integral form of the two-dimensional incompressible Navier–Stokes equations under the arbitrary Lagrangian–Eulerian (ALE) formulation is given as follows:

$$\int_{\partial\Omega(t)} \mathbf{u} \cdot \mathbf{n} dS = 0$$
$$\frac{\partial}{\partial t}\int_{\Omega(t)} \mathbf{u} d\Omega + \int_{\partial\Omega(t)} \mathbf{u}(\mathbf{u}-\mathbf{w}) \cdot \mathbf{n} dS = -\int_{\partial\Omega(t)} p\mathbf{n} dS + \int_{\partial\Omega(t)} \boldsymbol{\tau} \cdot \mathbf{n} dS,$$

(2.1)

where $\mathbf{u} = [u,v]^T$ is the fluid velocity, $\mathbf{w} = [u_{grid}, v_{grid}]^T$ is the mesh velocity; for pure flow simulations with stationary boundaries, $\mathbf{w} = 0$. The variable $p$ represents the pressure, and $\tau$ denotes the viscous stress tensor.

Under a conventional time-stepping framework, the spatially discretized governing equations can be written in the following semi-discrete form:

$$\frac{d\mathbf{q}}{dt} = -\mathbf{R}(\mathbf{q})$$

(2.2)

Here, $\mathbf{q}$ denotes the discretized flow variables, and $\mathbf{R}$ represents the residual operator corresponding to the discretized governing equations. In this work, a second-order implicit scheme is adopted for the physical time integration to ensure temporal accuracy. In addition, a pseudo-time $\tau$ is introduced, independent of the physical time $t$ and Eq.(2.2) is rewritten as:

$$\frac{\mathbf{q}^{n+1,(k+1)} - \mathbf{q}^{n+1,(k)}}{\Delta\tau} + \frac{3\mathbf{q}^{n+1,(k+1)} - 4\mathbf{q}^n + \mathbf{q}^{n-1}}{2\Delta t} = -\mathbf{R}(\mathbf{q}^{n+1,(k)})$$

(2.3)

For Eq.(2.3), within each given physical time step, the inner pseudo-time iterations proceed until the residual satisfies a prescribed convergence criterion, yielding the numerical solution at that time level. This dual-time stepping implicit scheme can be interpreted as solving a steady problem within each physical time step through pseudo-time evolution, and its convergence behavior remains constrained by the dynamical properties of the original governing equations.

In the optimizing a discrete loss (ODIL) method, the spatially discretized governing equations are formulated as a residual minimization problem, with the objective function written as:

$$\mathcal{L}(q) = \frac{1}{2}\left\|\frac{dq}{dt} + R(q)\right\|_2^2 \tag{2.4}$$

In practice, flow fields at multiple time instants $q^{T_0 \to T_1} = [q^{T_0}, q^{T_0+1}, ..., q^{T_1-1}, q^{T_1}]^T$ can be optimized simultaneously. To maintain consistency with the time-stepping approach, the ODIL method in this work also adopts a second-order implicit scheme for the temporal discretization:

$$\mathcal{L}(q^{T_0 \to T_1}) = \frac{1}{2}\left\|\frac{3q^{T_0 \to T_1} - 4q^{T_0-1 \to T_1-1} + q^{T_0-2 \to T_1-2}}{2\Delta t} + R(q^{T_0 \to T_1})\right\|_2^2 \tag{2.5}$$

By solving for the minimum of this optimization problem, the flow solution that satisfies the discretized governing equations is obtained directly. The corresponding gradient descent update can be formally written as:

$$\frac{dq}{d\tau} = -\nabla_q \mathcal{L}(q), \tag{2.6}$$

where $\tau$ represents the optimization iteration parameters.

In this work, to ensure the accuracy and stability of the numerical solution, an O-type structured mesh is employed, centered on the cylinder, with the far-field boundary located at 20 cylinder diameters away. A schematic of the computational domain, mesh, and boundary conditions for the flow past a cylinder is shown in Fig. 1. To ensure consistency, the time-stepping method and the ODIL method use identical settings in terms of mesh, spatial discretization schemes, and boundary conditions. Specifically, the Navier–Stokes equations are discretized using the finite volume method. The velocity gradients are computed using a cell-based Green–Gauss approach, the convective terms in the momentum equations are discretized with a second-order upwind scheme, and the pressure terms are discretized using a second-order scheme. For cases involving mesh motion, the radial basis function (RBF) method is employed for dynamic mesh deformation.

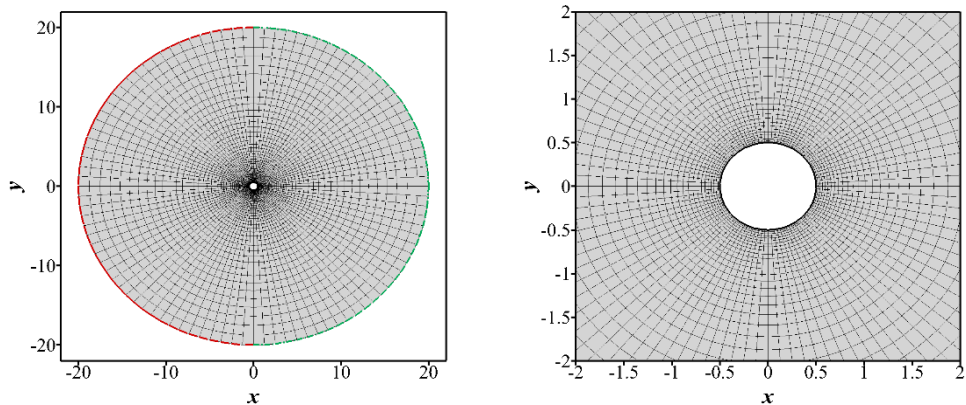

Fig. 1. Schematic illustration of the computational domain, mesh, and boundary conditions. The red line indicates

the velocity inlet boundary, the green line indicates the pressure outlet boundary, the black line indicates the wall boundary.

We first validated the accuracy and numerical consistency of the CFD solvers used in this work, based on both time-stepping and the ODIL framework. All solvers are implemented in Python, where pyGFSI_Iter and pyGFSI_Opt denote the results obtained from the time-stepping and ODIL approaches, respectively. In the ODIL computations, the L-BFGS optimizer provided in the *torch.nn* library is employed. Unless otherwise specified, the initial guess for the optimization is taken from the flow field at the previous physical time step, i.e., $q_{ini}^{T_0 \to T_1} = [q^{T_0-1}, q^{T_0-1}, ..., q^{T_0-1}]^T$. During each optimization process, the flow states over a time window of 200 physical time steps are optimized simultaneously. In addition, the residual tolerance of the governing equations is set to $10^{-8}$ in the optimization. For unsteady Navier–Stokes simulations, this tolerance is significantly stricter than the convergence criteria commonly used in conventional CFD solvers, ensuring that the obtained solutions satisfy the governing equations to a high degree in the discrete sense and thus possess strong numerical reliability.

Tables 1 and 2 present the computed amplitude of the lift coefficient $C_L^{max}$ and the dimensionless vortex shedding frequency $St$ for flow past a stationary cylinder at different Reynolds numbers. The Strouhal number $St$ is defined as $St = fD/U$, where $f$ is the vortex shedding frequency, $D$ is the cylinder diameter, and $U$ is the incoming flow velocity. It can be seen that the present results agree well with those reported in the literature, demonstrating the accuracy and reliability of the CFD methods based on both time-stepping and ODIL used in this work.

Table 1. Comparison of lift coefficient amplitudes $C_L^{max}$ at different Reynolds numbers

|  | 60 | 80 | 100 | 150 | 200 |
|---|---|---|---|---|---|
| **Baranyi[39]** | 0.130 | 0.240 | 0.320 | 0.510 | … |
| **Lu[40]** | 0.140 | 0.250 | 0.340 | 0.530 | 0.690 |
| **Le[41]** | … | 0.260 | … | … | 0.680 |
| **Wang[42]** | … | 0.260 | … | … | 0.710 |
| **pyGFSI_Iter** | 0.135 | 0.251 | 0.344 | 0.543 | 0.702 |
| **pyGFSI_Opt** | 0.135 | 0.249 | 0.344 | … | … |

Table 2. Comparison of Strouhal number $St$ under different Reynolds numbers

|  | 60 | 80 | 100 | 150 | 200 |
|---|---|---|---|---|---|
| **He[43]** | 0.135 | 0.153 | 0.167 | … | 0.198 |
| **Lu[40]** | 0.137 | 0.154 | 0.165 | 0.187 | 0.196 |
| **Wang[42]** | … | 0.158 | 0.170 | … | 0.195 |
| **Williamson[44]** | 0.136 | 0.152 | 0.164 | 0.179 | 0.183 |
| **pyGFSI_Iter** | 0.134 | 0.151 | 0.163 | 0.178 | 0.192 |
| **pyGFSI_Opt** | 0.136 | 0.151 | 0.162 | … | … |

Since this study focuses on the flow field induced by an oscillating cylinder, the force coefficients under forced motion are further validated. Following the work of Mittal et al. [45], the effect of oscillation frequency on the amplitude of the lift coefficient is examined at $Re = 33$, with the transverse oscillation amplitude fixed at $0.25D$. The computational setup in this work is consistent with that of Mittal et al., and the prescribed cylinder motion is given by: $h = 0.25D\sin(2\pi/U^*t)$, where $U^*$ denotes the reduced velocity. Fig. 2 shows the variation of the lift coefficient amplitude with the reduced velocity. The results are also compared with those obtained from the commercial software Fluent under identical mesh, spatial discretization schemes, and boundary conditions. The present results show excellent agreement with both Fluent and the results of Mittal et al., thereby confirming the accuracy and reliability of the CFD solvers based on time-stepping and ODIL method used in this work.

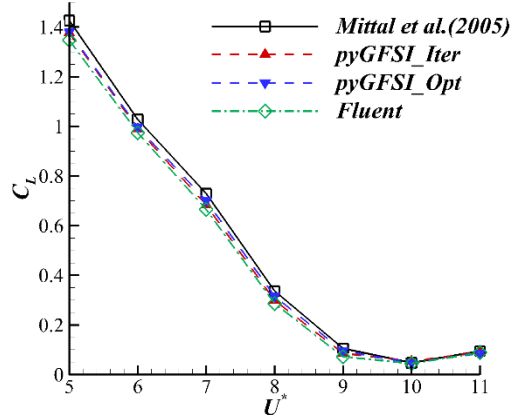

Fig. 2. Variation of the lift coefficient amplitude for a forced oscillating cylinder at $Re = 33$

## 2.2 Physics-informed neural networks and time-stepping-oriented neural network

In this section, we briefly introduce PINN and TSONN. A typical PINN employs a fully connected deep neural network (DNN) architecture to represent the solution $q(x,t)$ of the dynamical system. The network takes the spatial coordinates $x \in \Omega$ and time $t \in [0,T]$ as inputs and outputs the approximate solution $q(x,t;\theta)$. The network parameters $\theta$ are trained by minimizing a loss function that includes the residuals of the governing equations, as well as boundary and initial condition residuals. For the convenience of introducing TSONN, the PINN loss function is formulated as the mean squared error (MSE) of the residual vector:

$$\mathcal{L} = \frac{1}{N_0} \|f(q(\cdot;\theta))\|_2^2, \qquad (2.7)$$

where $N_0$ is the dimension of the residual vector $f(q)$. The residual vector $f(q)$ contains PDE residuals $g(q)$, boundary condition residuals $h(q)$ and initial condition residuals $i(q)$, and is weighed between different components through appropriate relative weights $\lambda_{PDE}$, $\lambda_{BC}$ and $\lambda_{IC}$, expressed as follows:

$$f(q) = \begin{bmatrix} \lambda_{PDE} g(q) \\ \lambda_{BC} \sqrt{N_g / N_h} h(q) \\ \lambda_{IC} \sqrt{N_g / N_i} i(q) \end{bmatrix} = 0, \qquad (2.8)$$

where, $N_g$, $N_h$ and $N_i$ are the dimensions of PDE residuals, boundary condition residuals and initial condition residuals respectively. Use DNN to predict the solution vector $q$ at a series of $m$ configuration points $D = \{x_i, t_i\}_{i=1}^m$, calculate the residual vector $f(q)$ through automatic differentiation, and then minimize the loss function (2.7) until convergence.

In many cases, due to the ill-conditioning of PINNs, it is difficult to obtain accurate solutions. TSONN addresses this issue by decomposing the original equations into a sequence of implicit pseudo-time-stepping equations, thereby transforming the ill-conditioned optimization problem into a series of well-conditioned sub-optimization problems:

$$\begin{aligned} &PINNs : f(q) = 0 \\ &TSONN : f(q) - (q - q_n)/\Delta\tau = 0 \\ &q_n = q(\cdot;\theta_n), n = 0, 1, \cdots, N \end{aligned} \qquad (2.9)$$

We minimize the residual of Eq.(2.9), which is referred to as the inner iteration. Once sufficient convergence is achieved after $K$ iterations, we proceed to the next optimization step and update $q_n$ using the latest network output, referred to as the outer iteration. The L-BFGS optimizer is employed, with the maximum number of iterations in each optimization step set to $K$, which naturally realizes the inner iterations. At each outer iteration, the optimizer is reinitialized to accommodate changes in the loss function, while the collocation points are randomly resampled.

In addition, we adopt a volume-weighted PDE residual [46] to scale the contributions of densely clustered grid points near the wall. This technique has been shown to perform better for problems with non-uniform grid distributions. Accordingly, the term $g(q)$ in Eq.(2.8) is replaced by $\tilde{g}(q)$, such that:

$$\tilde{g}(q) = \frac{g(q)v}{\sqrt{\frac{1}{N_g} \sum_{i=1}^{N_g} v_i^2}} = \frac{g(q)v}{\|v\|_2 / \sqrt{N_g}}, \qquad (2.10)$$

where $v_i$ represents the volume occupied by the collocation point in the computational domain. The denominator in the formula represents the root mean square grid volume, which is used for normalization.

In practical implementation, all relevant state parameters of the flow around the cylinder are included in the model input. As shown in Fig. 3, the inputs to the parametric solver cover all conditions of the viscous flow around the cylinder, including the spatial coordinates in the computational domain, the Reynolds number $Re$, the forcing amplitude $A$, and the reduced velocity $U^*$. It is worth noting that time $t$ is not directly used as an input. Instead, it is normalized by the oscillation period $T$, and $t/T$ is used as the temporal input. The prescribed motion of the cylinder is given by: $h = A \cdot \sin(2\pi / U^* t)$.

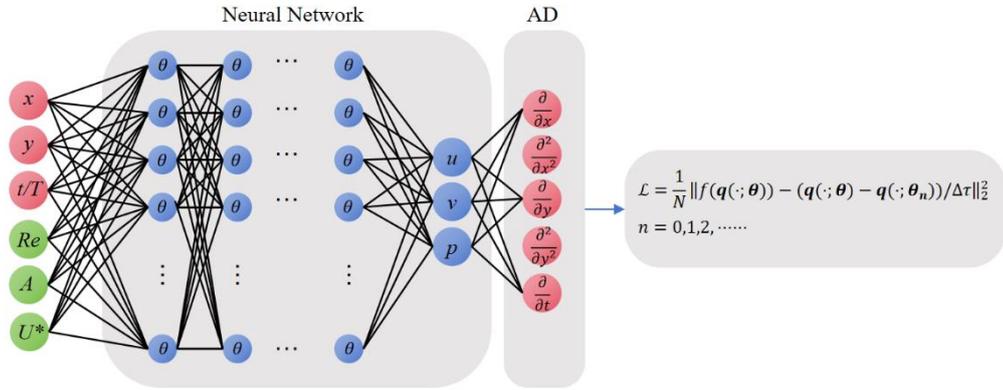

Fig. 3. Schematic diagram of parameterized solver based on TSONN

When solving the parametric problem, 30,000 interior collocation points, 2,000 boundary points and 2,000 initial points are randomly selected for each batch, together with randomly sampled flow conditions within the prescribed parameter space. These are then concatenated to form the input vectors. During training, for each collocation point $(x_i, y_i)$, given its corresponding time and flow state $((t/T)_i, Re_i, A_i, U_i^*)$, the displacement of the cylinder surface is first computed. The displacement at the far-field boundary is set to zero. Then, based on the physical distance from the point to the cylinder surface and the far-field boundary in the mesh, a weighted linear interpolation is performed to obtain the normal displacement at the collocation point. The deformed and updated coordinates $(\tilde{x}_i, \tilde{y}_i)$ are then combined with the corresponding flow state variables to form $(\tilde{x}_i, \tilde{y}_i, (t/T)_i, Re_i, A_i, U_i^*)$, which serves as the input to the neural network. This procedure ensures smooth and monotonic deformation, thereby preserving mesh regularity.

This study primarily focuses on using PINNs to identify different flow solution structures or modes and to provide reasonable initial guesses for the subsequent ODIL optimization, rather than employing PINNs as high-fidelity numerical solvers. Therefore, it is not required for the PINN to achieve accuracy comparable to traditional high-precision CFD solvers, nor is excessive training or strict residual convergence necessary. Specifically, the selected parameter ranges are as follows: Reynolds numbers from 5 to 100, forced oscillation amplitudes from 0.05 to 0.35, and reduced velocity from 7 to 13. This parameter space covers both forced flow regimes below the critical Reynolds number and multiple dynamical regimes above it, including flow states both inside and outside the conventional lock-in region. By training over a continuous parameter space, the network weights are shared across different parameter samples, which introduces a preference for structural continuity in the solution during optimization. Empirically, the network tends to approximate solutions along relatively continuous branches in parameter space, rather than frequently switching between flow modes with significantly different topological structures. This implicit structure-preserving property facilitates the identification of flow solutions with consistent modal characteristics over a wide parameter range and provides stable initial guesses for subsequent ODIL-based refinement.

In addition, the PINN training in this work employs the L-BFGS optimizer with a default learning rate of 1.0. The number of inner iterations $K$ (i.e., the maximum number of L-BFGS iterations in each outer iteration) is set to 500, and the history size of the optimizer is also set to 500. The relative weight is taken as $\lambda_{PDE}=100$, $\lambda_{BC}=\lambda_{IC}=1$, and the pseudo-time step is set to $\Delta\tau=0.3$. These parameters are selected based on numerical experience and have been validated in related studies [30, 31]. For the network architecture, a deep neural network with 6 hidden layers is used, each containing 256 neurons, with the hyperbolic tangent (tanh) activation function. All simulations are initialized using the flow field obtained at $Re=33$, and the maximum normalized time is set to $(t/T)_{max}=6$, corresponding to six oscillation periods of the forced motion. According to numerical experience, the flow typically evolves to a relatively stable periodic state within this time scale. Therefore, the flow field in the final period is regarded as an approximate periodic solution and is used as the initial condition for subsequent ODIL optimization and further computations.

## 3 Numerical Results

In this chapter, the periodic flow field obtained by PINNs is used both as the initial condition and as the initial guess for the optimization. Continuation simulations are then performed using both the conventional time-stepping method and the ODIL framework, in order to compare the differences in the resulting flow evolution. To ensure consistent temporal resolution, each oscillation period is discretized into $N_t=50$ physical time steps. In the ODIL optimization process, we jointly optimize the flow field over 200 physical time steps simultaneously, which covers 4 complete forced oscillation cycles. Specifically, the single-period approximate periodic solution obtained from PINNs is replicated four times in time to construct a continuous four-period flow sequence, which is used as the initial guess for ODIL.

First, to verify the consistency of different numerical methods in capturing the physical attracting state, a representative lock-in case ($Re = 80$, $A = 0.25$, $U^* = 7.5$) is considered. This case lies within the classical lock-in regime, where the vortex shedding frequency is phase-locked with the structural oscillation frequency. The corresponding results are shown in Fig. 4.

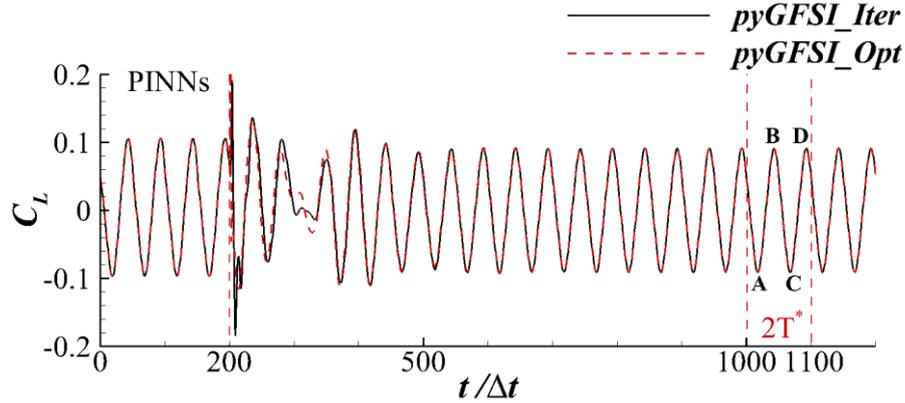

Fig. 4. Time history of the lift coefficient $C_L$ at $Re = 80$, $A = 0.25$, and $U^* = 7.5$ (within the lock-in regime).

Fig. 4 shows the time evolution of the lift coefficient $C_L$. The horizontal axis represents the physical time steps, expressed as $t/\Delta t$, where $\Delta t$ denotes the physical time increment, defined as $\Delta t = T/N_t$, allowing a clearer visualization of the periodic relationship between the imposed oscillation and the flow response. The first 200 time steps correspond to the periodic solution directly predicted by the PINN. This solution is then used as the initial condition for continuation using both the conventional time-stepping method and the ODIL framework, in order to assess its dynamical stability and numerical consistency. It can be observed that, during the continuation process, the lift signals obtained by both methods remain strictly periodic, with no noticeable drift or modulation in amplitude or frequency. The periodic solution obtained by the PINN remains stable under time integration, indicating that it not only satisfies the governing equations in the residual sense, but also corresponds to an attracting state in the dynamical system sense.

To facilitate the analysis of the spatiotemporal evolution of the flow structures, two representative oscillation periods are highlighted, and four characteristic instants, labeled A–D, are selected. The instantaneous vorticity fields at these instants are shown in Fig. 5. It can be observed that, under lock-in conditions, the wake exhibits a regular Bénard–von Kármán vortex street with alternating vortex shedding. The shedding pattern follows the classical 2S mode, in which a single vortex is shed from each side of the cylinder during one oscillation period. As the cylinder oscillates, the near-wake shear layers roll up periodically and are convected downstream, forming a highly symmetric and spatially periodic arrangement of positive and negative vortices. The four instants A–D correspond to different phase positions over two complete oscillation periods and clearly illustrate the processes of vortex formation, intensification, and shedding. This strict periodicity and the sustained 2S mode further confirm a phase-locked relationship between the vortex shedding frequency and the structural oscillation frequency.

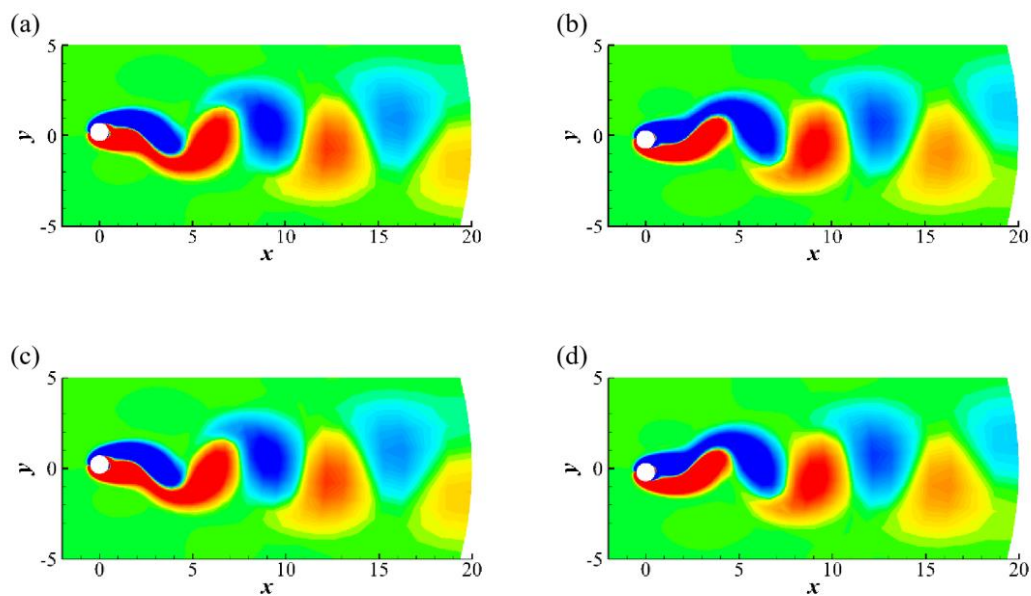

Fig. 5. Instantaneous vorticity fields at selected instants: (a) time-stepping method at instants A and C; (b) time-stepping method at instants B and D; (c) optimization-based method at instants A and C; (d) optimization-based method at instants B and D.

Once the flow reaches a stable periodic state, the force–displacement phase portrait of the lift coefficient is plotted, as shown in Fig. 6. The phase portrait exhibits a smooth and closed limit cycle, with no secondary loops or thickness spreading. The limit cycle trajectories obtained from different numerical methods overlap closely, indicating that only a single attracting periodic solution exists within the lock-in regime. This closed trajectory also confirms the establishment of a stable phase-locking relationship between the flow and the structural oscillation, with the oscillation frequency fully governed by the forcing frequency.

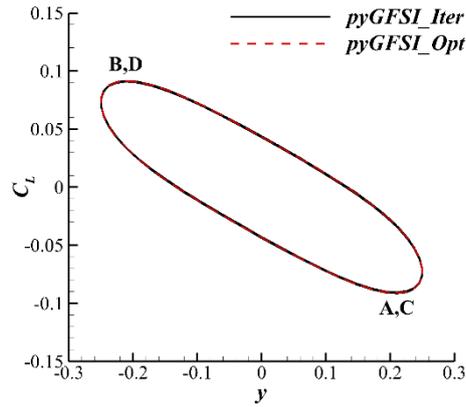

Fig. 6. Force–displacement phase portrait of the lift coefficient $C_L$ at $Re$ = 80, $A$ = 0.25, and $U^*$ = 7.5 (within the lock-in regime)

In summary, within the lock-in regime, both the conventional time-stepping method and the ODIL framework converge to the same stable periodic limit cycle, with no essential differences in the numerical results. This demonstrates the consistency of different numerical strategies in capturing the long-time dynamical behavior when a physical attracting state exists.

To further investigate the dynamical characteristics of the wake outside the lock-in regime, a representative case with reduced velocity $U^*$=10 ($Re$=80, $A$=0.25) is selected for analysis, as shown in Fig. 7. This case is chosen because, at Re=80, the natural vortex shedding frequency of a stationary cylinder corresponds to a Strouhal number of approximately $St \approx 0.15$. Under this condition, the imposed oscillation frequency of the structure is close to a 2:3 ratio with the natural shedding frequency. Therefore, this case provides a representative example for examining the interaction between natural vortex shedding and external forcing.

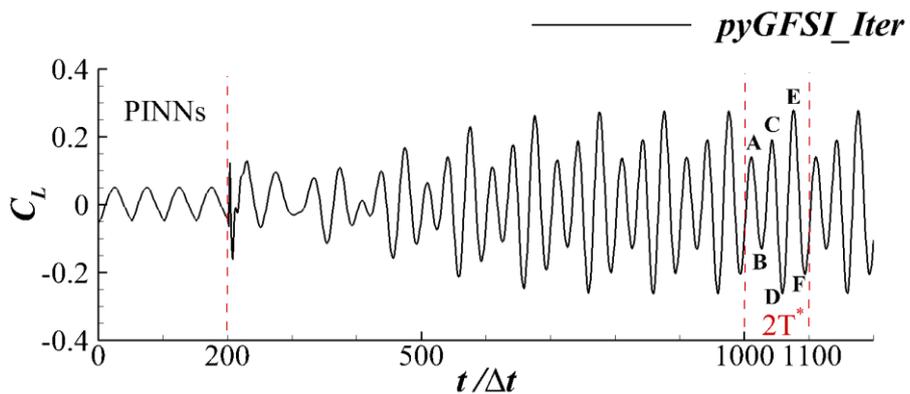

Fig. 7. Time history of the lift coefficient $C_L$ at $Re$ = 80, $A$ = 0.25, and $U^*$ = 10 (outside the lock-in regime) obtained using the time-stepping method.

Fig. 7 shows the time history of the lift coefficient $C_L$. After the initial transient, unlike the single-frequency periodic response observed within the lock-in regime, the lift coefficient here exhibits a clear multi-periodic structure. Specifically, in this stage, the signal repeats over a cycle spanning two oscillation periods, within which six pronounced positive and negative peaks can be identified. These peaks are labeled A–F in the figure. This behavior indicates that the wake dynamics are no longer governed by a single frequency, but instead reflect the combined influence of the natural vortex shedding frequency and the external forcing frequency.

To further quantify the frequency components of the wake oscillation under this condition, a Fourier transform is performed on the lift coefficient time series. The resulting spectrum is shown in Fig. 8, where the horizontal axis represents frequency and the vertical axis denotes the amplitude of the corresponding frequency components. It can be observed that two distinct dominant peaks appear in the spectrum obtained from the time-stepping solution, located at 0.02 Hz and 0.03 Hz, respectively. In the present numerical setup, one oscillation period contains 50 time steps. For the stationary cylinder flow at $Re = 80$, the Strouhal number is approximately $St \approx 0.15$, such that 0.03 Hz corresponds to the natural vortex shedding frequency of the wake. The other dominant frequency at 0.02 Hz matches the imposed oscillation frequency of the structure. These results indicate that the wake dynamics are influenced simultaneously by the natural vortex shedding instability and the external forcing. As a result, the flow response contains both frequency components, which explains the multi-periodic behavior observed in the lift coefficient time history: the signal repeats over two oscillation periods and exhibits multiple peaks within each cycle.

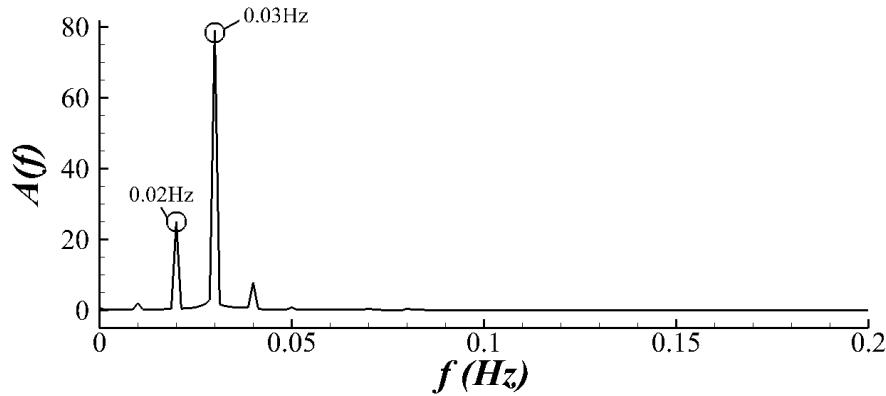

Fig. 8. Frequency spectrum of the lift coefficient time series obtained using the iterative solver for the case $Re$ = 80, $A$ = 0.25, $U^*$ = 10 (outside the lock-in regime).

Fig. 9 presents the force–displacement phase portrait of the lift coefficient after the flow reaches a stable oscillatory state. It can be observed that the trajectory is no longer a simple closed elliptical curve associated with single-frequency oscillations, but instead exhibits a star-like multi-lobed structure. The trajectory fully repeats itself after two vibration cycles, which is consistent with the approximate 2:3 ratio between the two characteristic frequencies in the system. Owing to this rational relationship between the natural vortex shedding frequency and the imposed oscillation frequency, the minimum repeating period of the system response corresponds to two forcing cycles, leading to a periodic orbit with multiple turning points in phase space.

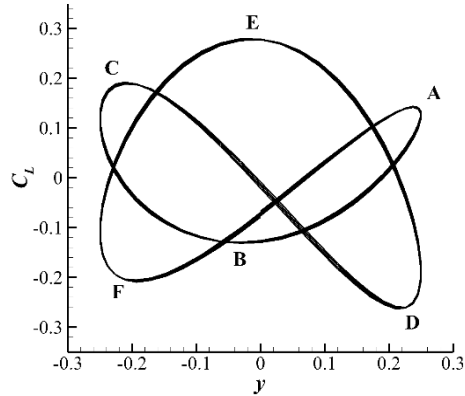

Fig. 9. Force–displacement phase portrait of the lift coefficient $C_L$ at $Re = 80$, $A = 0.25$, and $U^* = 10$ (outside the lock-in regime) obtained using the time-stepping method.

The instantaneous vorticity fields corresponding to six representative instants (A–F) in Fig. 7 are shown in Fig. 10. It is evident that the wake structure under this condition differs significantly from the classical 2S mode observed in the lock-in regime. In the standard 2S mode, one shear-layer vortex is typically generated during each upward and downward motion of the cylinder, resulting in two alternately shed single vortices per vibration cycle. In contrast, the present case exhibits pronounced asymmetry in the vortex shedding process. Over two forcing cycles, the vortex formation can be summarized as follows: during the first cycle, one single vortex is generated in both the upward and downward motions; during the second cycle, the shear-layer roll-up in the upward motion is significantly intensified, producing not only one primary vortex but also an additional pair of counter-rotating vortices, while a single vortex is still generated during the subsequent downward motion.

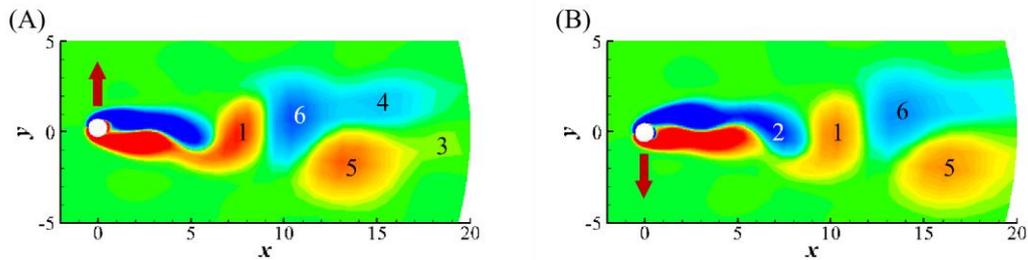

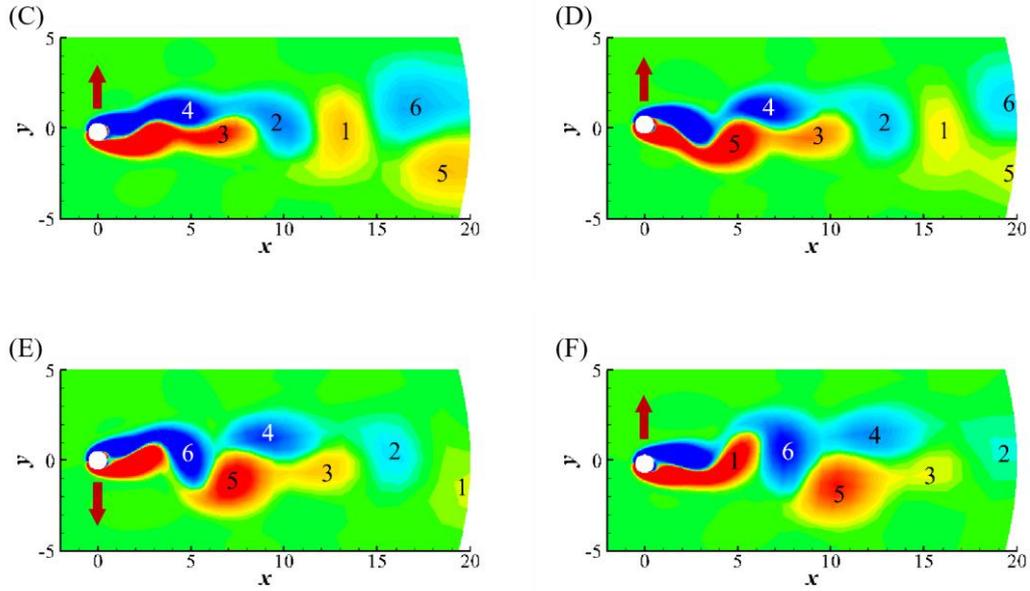

Fig. 10. Instantaneous vorticity fields at representative instants A–F obtained using the iterative solver for the case $Re = 80$, $A = 0.25$, $U^* = 10$ (outside the lock-in regime).

Therefore, within two vibration cycles, a total of six vortex structures are shed in the wake, forming a complex pattern characterized by the alternating appearance of single vortices and vortex pairs. Since this structure consists of a combination of single vortices (S) and vortex pairs (P), it can be classified as a generalized P+S vortex shedding mode. When the flow is outside the lock-in regime, the inherent wake instability persists and interacts with the external oscillation. This coupling leads to the generation of additional vortex pairs during the shear-layer roll-up process. This mechanism explains the composite periodic behavior observed in the time history, as well as the presence of multiple lift peaks within two vibration cycles.

Next, the flow solution obtained from the optimization-based method is analyzed, and the same dynamical characteristics as discussed above are examined. Fig. 11 presents the time history of the lift coefficient $C_L$ obtained from the optimization. It can be observed that, after the initial transient, the lift coefficient reaches a stable periodic state whose period exactly matches the imposed oscillation period of the structure. In contrast to the multi-periodic behavior observed in the time-stepping results, the present signal exhibits a strictly single-period repeating pattern, indicating that the wake dynamics are fully governed by the imposed oscillation.

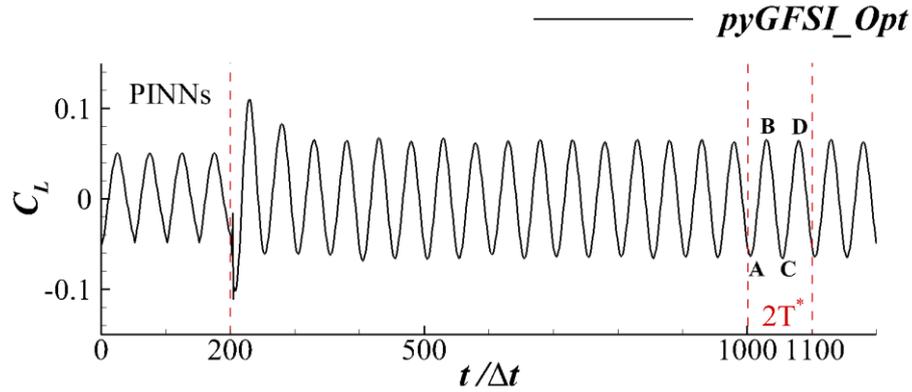

Fig. 11. Time history of the lift coefficient $C_L$ at $Re = 80$, $A = 0.25$, and $U^* = 10$ (outside the lock-in regime) obtained using the optimization-based method.

The corresponding force–displacement phase portrait is shown in Fig. 12. It can be observed that the trajectory forms a smooth and stable closed curve, with successive cycles collapsing onto a single path. The multi-lobed, star-like structure seen in the time-stepping results disappears, and the phase portrait becomes similar to that observed within the lock-in regime. This indicates that the system response has transitioned from a dual-frequency coupled state to a single-frequency dominated periodic motion, with a stable phase-locked relationship re-established between the flow and the structural oscillation.

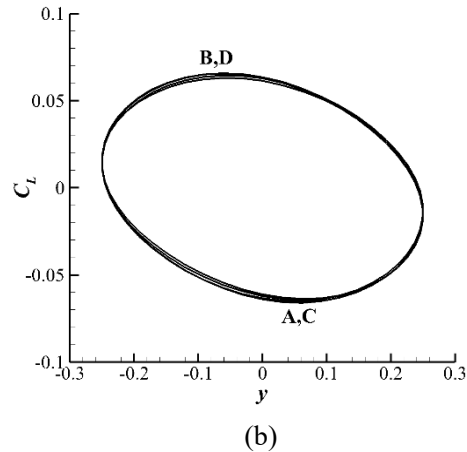

(b)

Fig. 12. Force–displacement phase portrait of the lift coefficient $C_L$ at $Re = 80$, $A = 0.25$, and $U^* = 10$ (outside the lock-in regime) obtained using the optimization-based method.

To further quantify the frequency composition, a Fourier transform of the lift coefficient signal is performed, and the resulting spectrum is shown in Fig. 13. It can be seen that the optimized solution exhibits a single dominant peak at 0.02 Hz, which corresponds exactly to the imposed oscillation frequency. In contrast, the time-stepping results also contain another primary peak at 0.03 Hz, associated with the natural vortex shedding frequency under this condition. The absence of the natural shedding frequency in the optimized solution indicates that the flow is governed solely by the imposed oscillation, containing only a single-frequency component.

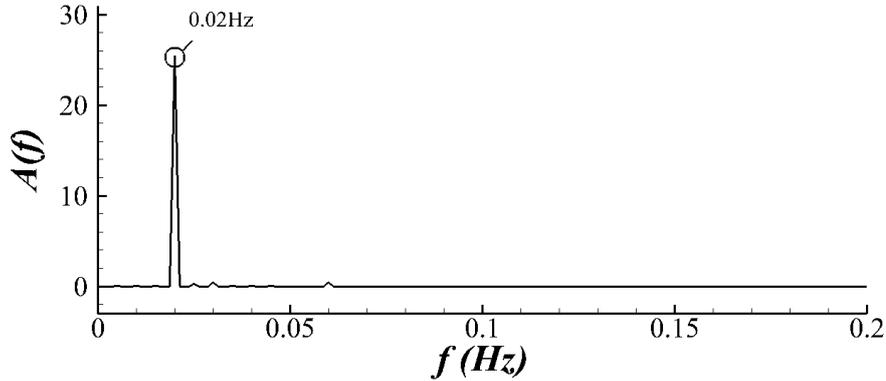

Fig. 13. Frequency spectrum of the lift coefficient time series obtained using the optimization-based solver for the case $Re = 80$, $A = 0.25$, and $U^* = 10$ (outside the lock-in regime).

The instantaneous vorticity fields at representative instants A–D in Fig. 11 are shown in Fig. 14. The vorticity contours clearly reveal that the wake exhibits a typical 2S vortex shedding mode. In this mode, during each half oscillation cycle, a single vortex is generated and shed from one side of the cylinder. As the upper and lower shear layers act alternately, one pair of oppositely signed single vortices is formed over a complete oscillation period. This wake structure is essentially identical to that observed within the lock-in regime.

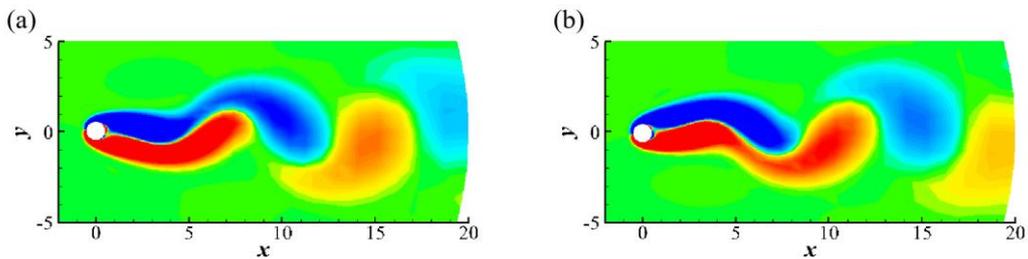

Fig. 14. Instantaneous vorticity fields at selected instants: (a) at instants A and C; (b) at instants B and D.

A comprehensive analysis of the time history, phase portrait, frequency spectrum, and vorticity fields shows that the flow solution obtained by the optimization corresponds to a single-period state entirely governed by the imposed oscillation frequency. This solution exhibits strictly periodic behavior in time, a stable closed trajectory in phase space, a wake structure characterized by the classical 2S vortex shedding mode, and a spectrum containing only a single dominant peak at the forcing frequency. However, it should be emphasized that, under the present parameter conditions, this single-period solution is not an attracting state of the original dynamical system. In conventional time-stepping simulations, the system naturally evolves toward a dual-frequency oscillatory state that contains both the natural vortex shedding frequency and the forcing frequency, rather than remaining on this single-frequency periodic solution. Therefore, this solution should be regarded as a non-attracting periodic state in the dynamical system sense.

Although this periodic solution is non-attracting, it can still be obtained by the optimization initialized by the PINN-predicted flow field. Since the solution satisfies both the governing equations and the imposed periodicity constraints, it confirms the existence of this flow state as a valid solution of the system.

This behavior is analogous to unstable steady solutions in fluid mechanics: although such steady states cannot be sustained under time integration, they still satisfy the governing equations and play an important role in the analysis of flow bifurcation and dynamical structure. From this perspective, the periodic solution identified in this work can be interpreted as a counterpart of unstable steady solutions under forced oscillation. Its existence suggests that, outside the lock-in regime, in addition to the dual-frequency oscillatory state obtained through time-stepping, there may also exist a branch of single-frequency periodic solutions. These branches, however, are dynamically unstable and therefore remain hidden to conventional simulation approaches.

To further assess the numerical reliability and robustness of the obtained periodic solution, additional tests on grid independence and sensitivity to initial conditions under this case are provided in the Appendix A. Specifically, the optimization is repeated on an alternative non-structured computational mesh, while simultaneously introducing significant perturbations to the PINN-generated initial field. Despite these modifications, the solution converges to a nearly identical periodic state, with consistent flow structures, dominant frequency, and force histories, indicating good robustness. Detailed results are provided in the Appendix A.

To assess the generality of the above findings, two additional cases, ($Re = 70$, $A = 0.25$, $U^* = 11$) and ($Re = 90$, $A = 0.25$, $U^* = 9$), are examined. In these cases, the imposed oscillation frequency is no longer in a simple rational ratio with the natural vortex shedding frequency of the cylinder, and therefore the system response does not exhibit clear periodic repetition. Instead, the dynamics become more complex. The time histories of the lift coefficient obtained from conventional time-stepping simulations are shown in Figs. 15 and 16.

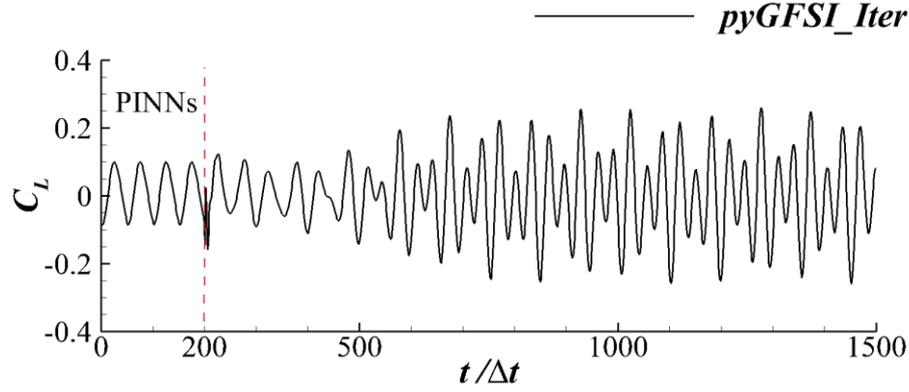

Fig. 15. Time history of the lift coefficient $C_L$ at $Re = 70$, $A = 0.25$, and $U^* = 11$ (outside the lock-in regime) obtained using the time-stepping method.

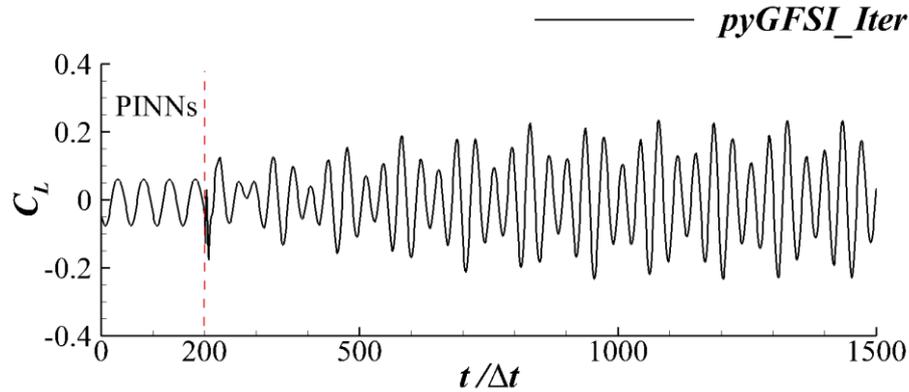

Fig. 16. Time history of the lift coefficient $C_L$ at $Re = 90$, $A = 0.25$, and $U^* = 9$ (outside the lock-in regime) obtained using the time-stepping method.

It can be observed that, the lift signals in both cases exhibit clear multi-frequency oscillations. The waveforms do not repeat over long time intervals, indicating a non-periodic response. This suggests that the flow dynamics are simultaneously influenced by the imposed oscillation and the natural vortex shedding mechanism.

To further analyze the frequency content, a Fourier transform is applied to the lift coefficient time series, and the resulting spectra are shown in Figs. 17 and 18. The spectral results indicate the presence of two dominant frequency components in both cases. In particular, a peak at 0.02 Hz appears in both cases, corresponding to the imposed oscillation frequency. The additional peak at 0.03125 Hz in Fig. 17 corresponds to the natural vortex shedding frequency at $Re = 70$ ($St \approx 0.141$), while the peak at 0.028 Hz in Fig. 18 corresponds to the natural shedding frequency at $Re = 90$ ($St \approx 0.157$). These results confirm that, in both cases, the flow response is governed by the combined effects of structural forcing and natural wake instability. Since the two frequencies are not in a simple rational ratio, the lift signal does not exhibit periodic repetition.

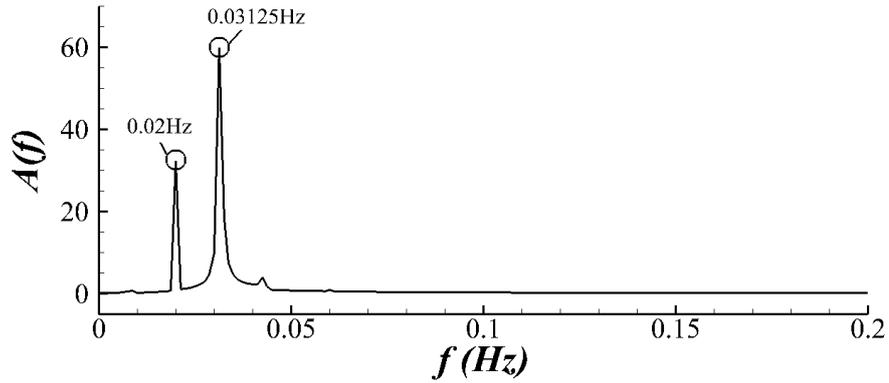

Fig. 17. Frequency spectrum of the lift coefficient time series obtained using the iterative solver for the case $Re = 70$, $A = 0.25$, and $U^* = 11$ (outside the lock-in regime).

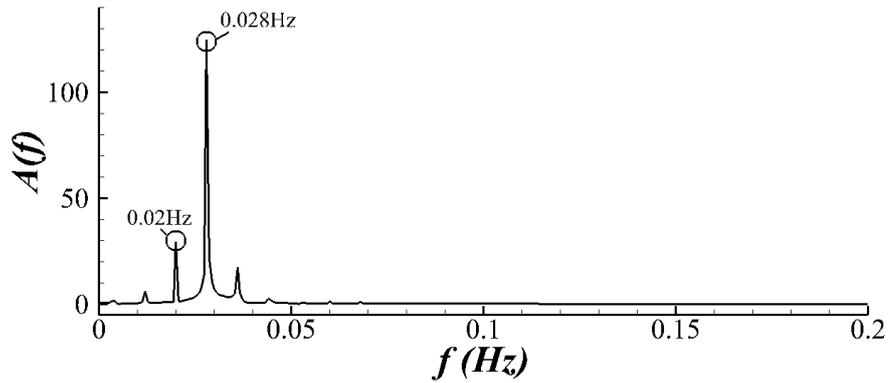

Fig. 18. Frequency spectrum of the lift coefficient time series obtained using the iterative solver for the case $Re = 90$, $A = 0.25$, and $U^* = 9$ (outside the lock-in regime).

Subsequently, the same cases are solved using the ODIL framework. The resulting time histories of the lift coefficient are shown in Figs. 19 and 20. It can be observed that the oscillation behavior differs markedly from that obtained by time-stepping. In the optimization results, the lift signal exhibits a stable periodic oscillation with a period identical to the imposed oscillation, indicating that the system response is fully synchronized with the external forcing.

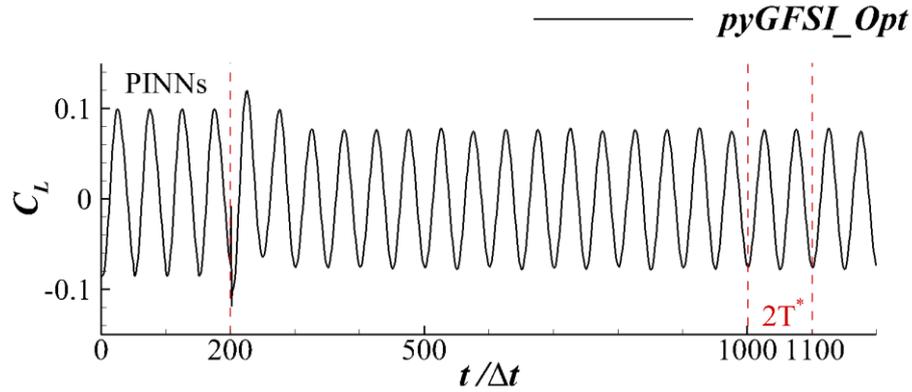

Fig. 19. Time history of the lift coefficient $C_L$ at $Re = 70$, $A = 0.25$, and $U^* = 11$ (outside the lock-in regime) obtained using the optimization-based method.

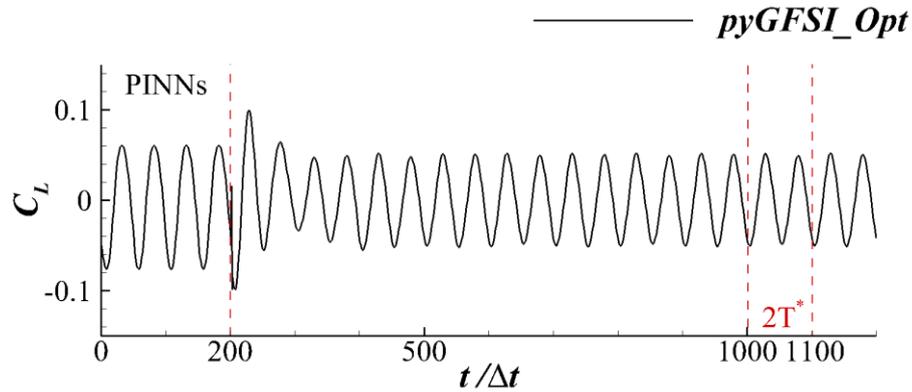

Fig. 20. Time history of the lift coefficient $C_L$ at $Re = 90$, $A = 0.25$, and $U^* = 9$ (outside the lock-in regime) obtained using the optimization-based method.

A Fourier analysis of the optimized results is further performed to confirm this behavior, with the spectra shown in Figs. 21 and 22. The spectral results indicate that, in both cases, only a single dominant frequency at 0.02 Hz—corresponding to the imposed oscillation—is present, while the frequency component associated with natural vortex shedding disappears. This demonstrates that the optimization-based approach is able to obtained single-frequency periodic solutions governed solely by the imposed oscillation. This observation is consistent with the previous cases and further confirms that the optimization method captures the corresponding non-attracting periodic solutions.

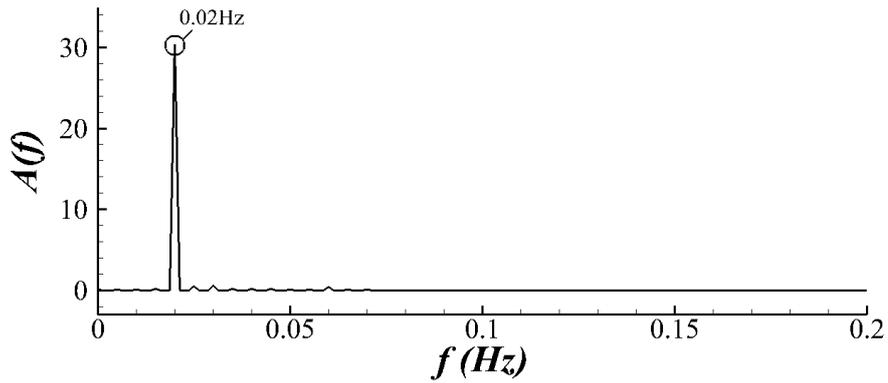

Fig. 21. Frequency spectrum of the lift coefficient time series obtained using the optimization-based solver for the case $Re = 70$, $A = 0.25$, and $U^* = 11$ (outside the lock-in regime).

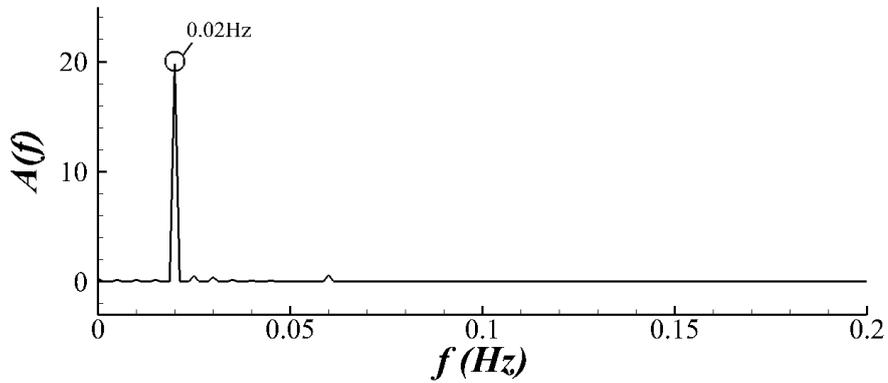

Fig. 22. Frequency spectrum of the lift coefficient time series obtained using the optimization-based solver for the case $Re = 90$, $A = 0.25$, and $U^* = 9$ (outside the lock-in regime).

The above results indicate that, outside the lock-in regime, in addition to the dual-frequency oscillatory solutions obtained by conventional time-stepping methods, there also exist single-frequency periodic solutions that satisfy the governing equations but are non-attracting. Moreover, different numerical approaches may converge to different types of solutions during the computation.

To further understand the origin of this phenomenon, the next chapter examines the problem from the perspective of numerical evolution mechanisms. A systematic comparison between conventional time-stepping methods and optimization-based approaches is conducted, with particular emphasis on their convergence pathways in the solution space and the associated dynamical implications. This analysis aims to elucidate why optimization-based methods are able to capture non-attracting periodic solutions that are difficult to obtain using standard time-stepping techniques.

# 4 Numerical Evolution Mechanism Analysis

## 4.1 Linear Stability Analysis of the Steady Problem

First consider the general form of the steady governing equation:

$$f(q) = 0, \tag{4.1}$$

where $q$ represents the discretized flow field variable vector, and $f(q)$ is the residual vector corresponding to the discrete governing equation. Linearizing the system around an equilibrium solution yields:

$$f(q) = Aq - b = 0, \tag{4.2}$$

where the matrix $A$ is the Jacobian matrix of the governing equations evaluated at the equilibrium point.

### 4.1.1 Solving Steady Problems Using Time-Stepping Methods

For conventional time-stepping methods, the basic idea is to construct a pseudo-time evolution process that drives the system toward the solution of the governing equations. An explicit time-stepping scheme can be written as:

$$\frac{q^{(k+1)} - q^{(k)}}{\Delta \tau} = -f(q^{(k)}), \tag{4.3}$$

where $\Delta \tau$ denotes the pseudo-time step. Substituting the linearized expression yields:

$$q^{(k+1)} = q^{(k)} - \eta(Aq^{(k)} - b), \tag{4.4}$$

where $\eta = \Delta \tau$.

It can thus be seen that the time-stepping algorithm essentially approaches the solution of Eq. (4.2) through an iterative process. In the continuous limit, this process can be described by the following pseudo-dynamical system:

$$\dot{q} = -(Aq - b) \tag{4.5}$$

The stability of this dynamical system near an equilibrium point is determined by the eigenvalues of matrix $A$. If all eigenvalues of $A$ are positive, the equilibrium point is stable, and the time-stepping procedure converges toward this solution. In contrast, if $A$ has negative eigenvalues, unstable modes exist along the corresponding directions, and small perturbations will grow during the time evolution, causing the solution to deviate from this state and eventually evolve toward other stable states. Therefore, for time-stepping methods, numerical simulations can typically converge only to attracting solutions of the original dynamical system.

### 4.1.2 Solving Steady Problems Using Optimization-Based Methods

For optimization methods, a quadratic residual loss function is typically constructed as:

$$\mathcal{L} = f^T f = (Aq - b)^T (Aq - b) \tag{4.6}$$

Taking the gradient with respect to $q$ yields:

$$\nabla_q \mathcal{L} = A^T (Aq - b) \tag{4.7}$$

Accordingly, the gradient descent update can be written as:

$$q^{(k+1)} = q^{(k)} - \eta A^T (Aq^{(k+1)} - b), \tag{4.8}$$

where $\eta$ denotes the optimization step size. It can be seen that this process is equivalent to solving the following normal equation:

$$A^T A q = A^T q \tag{4.9}$$

In the continuous limit, it corresponds to the following pseudo-dynamical system:

$$\dot{q} = -A^T (Aq - b) \tag{4.10}$$

The stability of this dynamical system near an equilibrium point is determined by the eigenvalues of the corresponding matrix $A^T A$. Since this matrix is positive semi-definite, all its eigenvalues are non-negative. Therefore, any solution that satisfies Eq. (4.2) can serve as a stable equilibrium solution of the optimization dynamics. This implies that even if a solution corresponds to an unstable equilibrium solution in the original dynamical system, the optimization algorithm may still converge to it. Consequently, compared with time-stepping methods, optimization-based approaches are capable of exploring multiple solution branches of the governing equations within a broader solution space.

### 4.1.3 Solving Steady Problems Using PINNs

For the Physics-Informed Neural Networks method under the numerical differentiation framework (PINNs-ND):

$$q = J(\phi), \tag{4.11}$$

where $\phi$ denotes the network parameters, and $J(\phi)$ represents the Jacobian of the network output $q$ with respect to the parameters $\phi$.

Substituting Eq. (4.11) into the residual function (4.6) yields the loss function:

$$\mathcal{L} = (AJ\phi - b)^T (AJ\phi - b) \tag{4.12}$$

Taking the gradient with respect to the network parameters $\phi$ gives:

$$\nabla_\phi \mathcal{L} = J^T A^T (AJ\phi - b) \tag{4.13}$$

Accordingly, the gradient descent update can be written as:

$$\phi^{(k+1)} = \phi^{(k)} - \eta J^T A^T (AJ\phi^{(k)} - b) \tag{4.14}$$

The corresponding pseudo-dynamical system is:

$$\dot{\phi} = -J^T A^T (AJ\phi - b) \tag{4.15}$$

It can be seen that the PINN method is essentially still a residual minimization problem, and its numerical evolution mechanism shares a similar structure with the aforementioned optimization-based methods. The stability of this dynamical system near an equilibrium point is governed by the eigenvalues of the matrix $J^T A^T AJ$, which is also positive semi-definite. Therefore, as long as the network-represented solution satisfies the governing equations, it can serve as a stable point of the optimization process, regardless of its stability in the original dynamical system.

## 4.2 Extension to Unsteady Problems

For unsteady problems, the governing equations must be discretized in time. Taking an explicit scheme as an example, the basic form can be written as:

$$\frac{q^{(n)} - q^{(n-1)}}{\Delta t} = -(Aq^{(n-1)} - b) \tag{4.16}$$

For conventional time-stepping methods, the numerical evolution process still essentially corresponds to the following pseudo-dynamical system:

$$\dot{q} = -(Aq - b) \tag{4.17}$$

This system shares the same stability structure as the original dynamical system, and therefore the time-stepping procedure can ultimately converge only to attracting solutions of the original system.

For optimization-based approaches, all discrete variables over the entire time interval $T_0 \sim T_1$ are typically optimized simultaneously. Denoting the space–time variables in vector form as $q_{T_0 \to T_1}$, one obtains the global linear system:

$$B q_{T_0 \to T_1} = b, \tag{4.18}$$

where the matrix $B$ is the Jacobian that incorporates both the temporal discretization operator and the spatial operator $A$. Compared with the matrix $A$ in the steady case, $B$ reflects both the spatial discretization structure and the time-marching relations.

Accordingly, for unsteady problems, the gradient descent iteration takes the form:

$$q_{T_0 \to T_1}^{(k+1)} = q_{T_0 \to T_1}^{(k)} - \eta B^T (B q_{T_0 \to T_1}^{(k)} - b), \tag{4.19}$$

with the corresponding pseudo-dynamical system:

$$\dot{q}_{T_0 \to T_1} = -B^T(Bq_{T_0 \to T_1} - b) \quad (4.20)$$

Similar to the steady case, the stability of this dynamical system is governed by a positive semi-definite matrix $B^T B$. As long as a space–time solution $q_{T_0 \to T_1}$ satisfies the governing equations, it is a stable equilibrium point of the optimization dynamics, regardless of whether it is an attracting state of the original dynamical system.

Similarly, for the PINNs method under the numerical differentiation framework (PINNs-ND), the gradient descent iteration with respect to the network parameters $\phi$ can be written as:

$$\phi^{(k+1)} = \phi^{(k)} - \eta J^T B^T (BJ\phi^{(k)} - b), \quad (4.21)$$

and the corresponding pseudo-dynamical system is:

$$\dot{\phi} = -J^T B^T (BJ\phi - b) \quad (4.22)$$

As in the steady case, the stability of this system is controlled by a positive semi-definite matrix $J^T B^T BJ$. Therefore, any solution represented by the network that satisfies the governing equations becomes a stable equilibrium point of the optimization process, independent of its attractivity in the original dynamical system.

It should be noted that the above stability analysis of PINNs is carried out under the numerical differentiation framework. Under the automatic differentiation framework, the residual is typically treated as a function of the input variables $x$, rather than an explicit function of the network parameters $\phi$. As a result, it is difficult to directly obtain the Jacobian $A$ and $B$, and an identical linear stability analysis cannot be performed. Nevertheless, regardless of whether automatic differentiation or numerical differentiation is employed, the training of PINNs essentially involves minimizing a squared residual loss via gradient-based optimization, and the underlying numerical mechanism remains the same. Therefore, from the perspective of the optimization structure and its associated dynamical evolution, the above analysis still captures the essential behavior of the PINN method. To further support this argument, a canonical Hopf bifurcation problem is constructed in the Appendix B and solved using PINNs within the automatic differentiation framework. The results show that the method is likewise capable of identifying two types of solutions with distinct dynamical properties, thereby providing additional evidence that optimization-based approaches can capture non-attracting solutions.

In summary, there exists a fundamental difference between time-stepping methods and optimization-based approaches in their numerical evolution mechanisms. Time-stepping methods advance the solution through successive temporal integration of the governing equations, and their numerical evolution follows that of the original dynamical system. As a result, their convergence behavior is determined by the stability of the original system, and they can ultimately obtain only attracting solutions. In contrast, optimization-based methods solve the discretized governing equations by minimizing the squared residual. Their stability is governed by the structure of the associated normal equations, such that any solution satisfying the governing equations can serve as a stable point of the optimization process, regardless of its attractivity in the original dynamical system.

Therefore, outside the lock-in regime, even if certain periodic solutions are non-attracting in the original system, optimization algorithms can still converge to them given appropriate initial guesses. In the present study, by using the PINN-predicted solution as the initial guess and performing further optimization with the ODIL framework, a single-period solution that satisfies the governing equations but is non-attracting has been successfully obtained for the forced oscillating cylinder at supercritical Reynolds numbers, which reveals the existence of additional solution structures in this system.

## 5 Conclusions and Outlook

This study investigates the structure of flow solutions in the forced oscillating cylinder problem at supercritical Reynolds numbers. Motivated by the classical Hopf bifurcation scenario in the flow past a stationary cylinder, where the system transitions from a steady state to a stable periodic vortex shedding state beyond a critical Reynolds number, the original steady solution—although still satisfying the governing equations—becomes inaccessible to time-stepping methods and is referred to as an unstable steady solution. Based on this understanding, the present work explores whether, in the forced oscillating cylinder flow at supercritical Reynolds numbers, there exist additional flow solutions that satisfy the governing equations but are non-attracting in the dynamical sense, beyond the attracting states obtained by conventional time-stepping methods.

In this study, when we employ Physics-Informed Neural Networks (PINNs) to solve the incompressible Navier–Stokes equations for the flow past a forced oscillating cylinder at supercritical Reynolds numbers, a class of flow solutions that cannot be accessed by direct time-stepping is discovered. These solutions are then used as initial guesses in the Optimizing a Discrete Loss (ODIL) framework to verify their existence. Subsequently, an analysis of the numerical evolution mechanisms is conducted to elucidate the properties of different solution branches and their accessibility. The main conclusions are summarized as follows:

1. For forced oscillating cylinder flows at supercritical Reynolds numbers, in addition to the attracting solutions obtained by conventional time-stepping methods, there exist single-frequency periodic solutions that satisfy both the governing equations and the oscillatory boundary conditions. These solutions contain only the imposed oscillation frequency and remain phase-locked with the cylinder motion, even when the corresponding parameters lie outside the conventional lock-in regime. In contrast, the flow responses obtained by time-stepping typically contain both the forcing frequency and the natural vortex shedding frequency.

2. By using the PINN-predicted solution as the initial guess in the ODIL framework, the solution can be stably maintained throughout the optimization process. This demonstrates that the solution strictly satisfies the governing equations and boundary conditions and is self-consistent in the optimization sense. In contrast, when using conventional time-stepping, the solution gradually departs from this state and eventually converges to a multi-frequency oscillatory flow containing the natural vortex shedding frequency, indicating that the single-frequency periodic solution is non-attracting in the original dynamical system.

3. From the perspective of numerical evolution mechanisms, the convergence behavior of time-stepping methods is constrained by the spectral properties of the Jacobian matrix $A$ obtained from the linearized governing equations, and thus they can only converge to attracting solutions of the dynamical system. In contrast, optimization-based methods minimize the squared residual, and their gradient evolution is governed by the associated normal matrix $A^T A$, which alters the numerical dynamical structure. As a result, solutions that satisfy the governing equations but are non-attracting in the dynamical sense can still be identified and maintained as minima of the optimization problem.

In summary, this study demonstrates that, in the forced oscillating cylinder flow, there exist non-attracting periodic solutions that satisfy the governing equations and boundary conditions, in addition to the attracting states obtained by conventional time-stepping methods. Optimization-based approaches provide a new numerical pathway for identifying and computing such solutions. The present results reveal the fundamental differences between time-stepping and optimization-based methods from both numerical and dynamical perspectives, and highlight the unique capability of optimization-based approaches in exploring multiple solution branches in complex flow systems. This work offers a new perspective for understanding the complex dynamics of forced oscillatory flows and provides a useful reference for future studies on non-attracting flow solutions using optimization-based methods.

## Appendix A

To further assess the numerical reliability and robustness of the obtained periodic solution, additional tests under the case (Re=80, A=0.25, U*=10) are performed by simultaneously modifying the computational mesh and perturbing the initial conditions. First, the computational grid is replaced by an unstructured C-type mesh, as shown in Fig. A1. Compared with the O-type structured mesh used in the main text, this mesh features a different topology and connectivity, allowing the sensitivity of the solution to spatial discretization to be evaluated. In addition, significant perturbations are introduced to the PINNs-generated initial guess to examine the sensitivity to initial conditions. Specifically, the flow field is projected onto a POD basis, and the corresponding modal coefficients are randomly perturbed by 30%. The perturbed flow field is then reconstructed and used as the initial condition for the subsequent optimization.

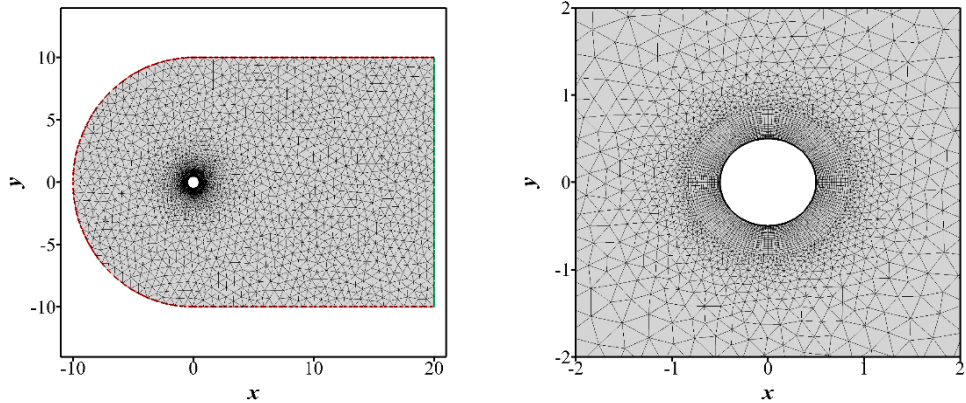

Fig. A1. Schematic illustration of the computational domain, mesh, and boundary conditions. The red line indicates the velocity inlet boundary, the green line indicates the pressure outlet boundary, the black line indicates the wall boundary.

The time history of the lift coefficient obtained from the ODIL optimization, along with its corresponding frequency spectrum, is shown in Figs. A2 and A3, respectively. Despite the combined modifications to both the computational mesh and the initial conditions, the solution still converges to the same periodic state as reported in the main text. After the initial transient, the lift coefficient exhibits a stable periodic oscillation, with the spectrum showing only a single dominant peak at 0.02 Hz corresponding to the imposed oscillation, while the component associated with natural vortex shedding is completely absent. The temporal evolution of the lift coefficient and the dominant frequency identified from the spectral analysis show excellent agreement with the results presented in the main text, indicating that the essential dynamical features of the solution are preserved under both mesh variation and initial perturbations.

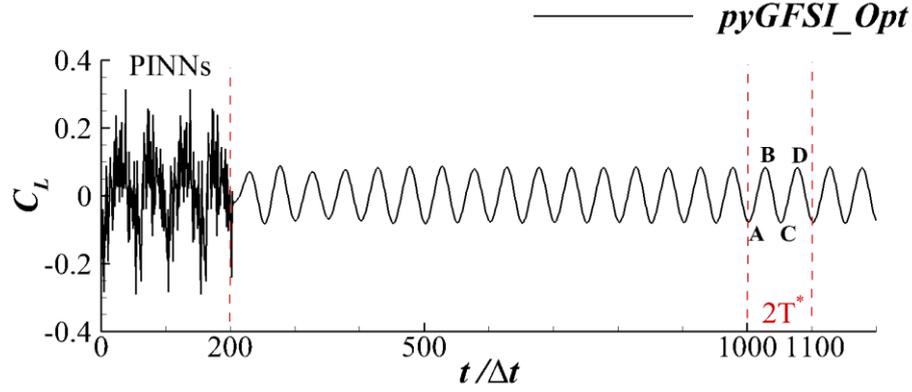

Fig. A2. Time history of the lift coefficient $C_L$ at $Re = 80$, $A = 0.25$, and $U^* = 10$ (outside the lock-in regime) obtained using the optimization-stepping method.

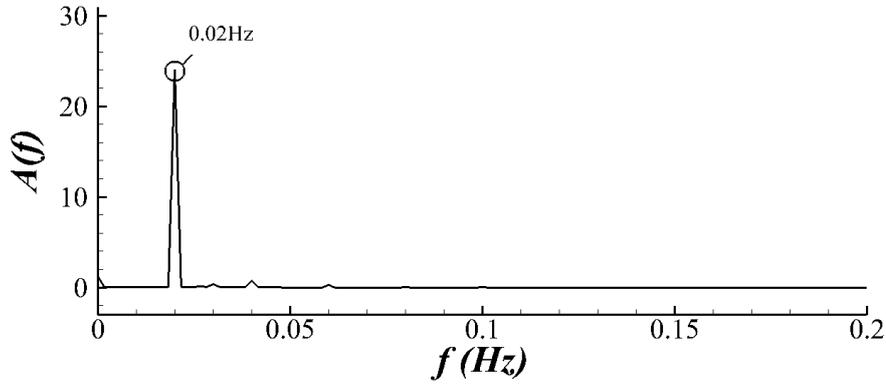

Fig. A3. Frequency spectrum of the lift coefficient time series obtained using the optimization-based solver for the case $Re = 80$, $A = 0.25$, and $U^* = 10$ (outside the lock-in regime).

The instantaneous vorticity fields at representative instants A–D in Fig. A2 are shown in Fig. A4. The wake structure again exhibits the characteristic 2S vortex shedding mode, with phase relationships and spatial organization consistent with those obtained in the main text. Compared with the results obtained in the main test, the vorticity contours in the present case appear slightly less smooth. This difference can be attributed to the use of the unstructured C-type mesh, which tends to introduce larger discretization errors and consequently results in reduced smoothness of the flow field, rather than indicating any fundamental change in the underlying flow physics.

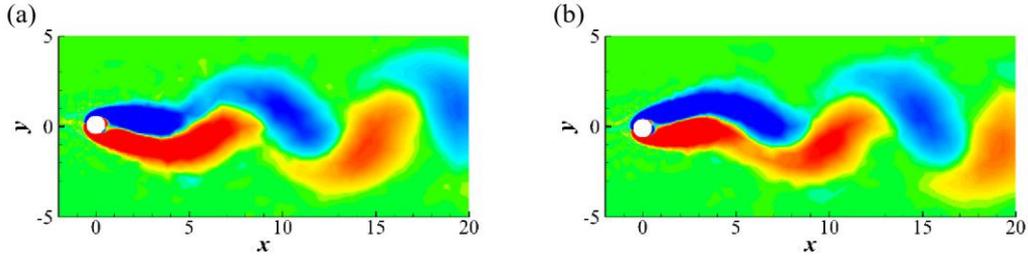

Fig. A4. Instantaneous vorticity fields at selected instants: (a) at instants A and C; (b) at instants B and D.

Overall, these results demonstrate that the periodic solution identified using PINNs and ODIL is robust with respect to variations in spatial discretization and substantial perturbations to the initial conditions. Although this solution is dynamically non-attracting in the original system, it can still be consistently recovered within the optimization framework. This suggests that the solution corresponds to a well-defined local minimum of the residual-based optimization problem, possessing a finite basin of attraction in the optimization landscape, even though it does not constitute an attracting state in the original dynamical system.

## Appendix B

To verify that physics-informed neural networks within the automatic differentiation framework (PINNs-AD) can obtain different types of solutions, a canonical Hopf bifurcation model is constructed in this section and solved numerically using PINNs-AD. This example is used to demonstrate that, under the same governing equations, the final outcome of the optimization process may depend on the initial parameters, leading to convergence toward different dynamical states.

Consider the following canonical Hopf bifurcation system:

$$\begin{aligned}\dot{x} &= \beta x - \omega y + \lambda x(x^2 + y^2) \\ \dot{y} &= \omega x + \beta y + \lambda y(x^2 + y^2),\end{aligned} \quad (B.1)$$

where $x(t)$ and $y(t)$ are the state variables, and $\beta, \omega$ and $\lambda$ are control parameters.

When $\beta < 0$, the system is stable at the origin $(0,0)$; when $\beta > 0$, the origin becomes an unstable equilibrium, and under perturbations the system moves away from it and eventually converges to a stable limit cycle. For $\lambda < 0$, the radius of the limit cycle is given by:

$$r = \sqrt{-\frac{\beta}{\lambda}} \quad (B.2)$$

In this study, the parameters are chosen as:

$$\omega = 1, \quad \lambda = -1, \quad \beta = 0.5 \quad (B.3)$$

Under these conditions, the origin is an unstable equilibrium and a stable limit cycle exists, with a theoretical radius of $r = \sqrt{0.5} \approx 0.707$. The system is expected to converge to this limit cycle during time evolution.

The above dynamical system is solved using TSONN. The neural network adopts a fully connected architecture with four hidden layers, each containing 16 neurons, and uses the hyperbolic tangent (tanh) activation function. The input to the network is the time variable $t$, while the outputs are the state variables $x(t)$ and $y(t)$. The maximum input time is set to $t_{max} = 30$, and 10,000 uniformly distributed sampling points are used over the time interval $t \in [0, 30]$ for training. The initial conditions are set to $x = 0$ and $y = 0$, with an associated weighting factor $\lambda_{IC} = 10$.

To investigate the effect of neural network initialization on the accessibility of different solution structures, numerical experiments are conducted by controlling the initialization range of the network weights. Specifically, the weights of the hidden layers are initialized using random values drawn from a uniform distribution within the interval $[-R_m, R_m]$, where $R_m$ is a scaling parameter controlling the initialization magnitude. Two different values of $R_m = 0.1$ and $R_m = 0.5$ are considered, and for each case, the network is randomly initialized four times and trained independently. This allows for a comparison of the convergence behavior of PINNs and the resulting solution structures under different initialization conditions.

To assess the nature and accuracy of the obtained solutions, two types of error metrics are defined. The first is the limit cycle error, used to measure the deviation of the solution from the theoretical limit cycle:

$$E_1 = x(t_{max})^2 + y(t_{max})^2 - r^2 \tag{B.4}$$

The second is the origin error, used to quantify whether the system remains near the origin:

$$E_2 = x(t_{max})^2 + y(t_{max})^2 \tag{B.5}$$

In the numerical experiments, the loss convergence histories and phase-space trajectories under different random initializations are plotted to compare the convergence behavior of PINNs, as shown in Figs. A5 and A6. Here, $R_m = 0.5$ corresponds to the $E_1$ error, while $R_m = 0.1$ initialization corresponds to the $E_2$ error.

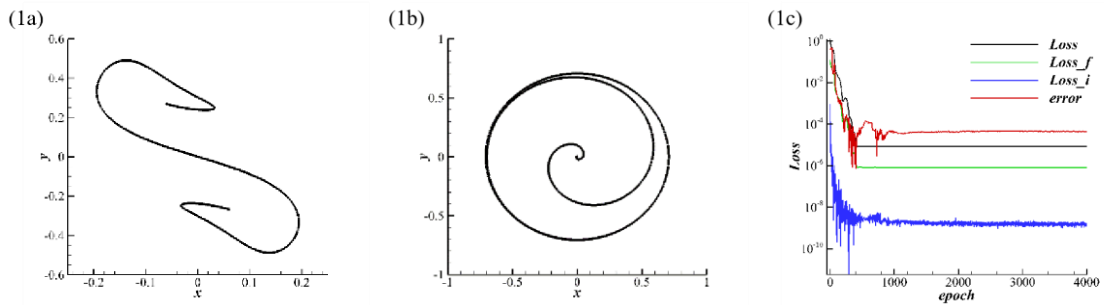

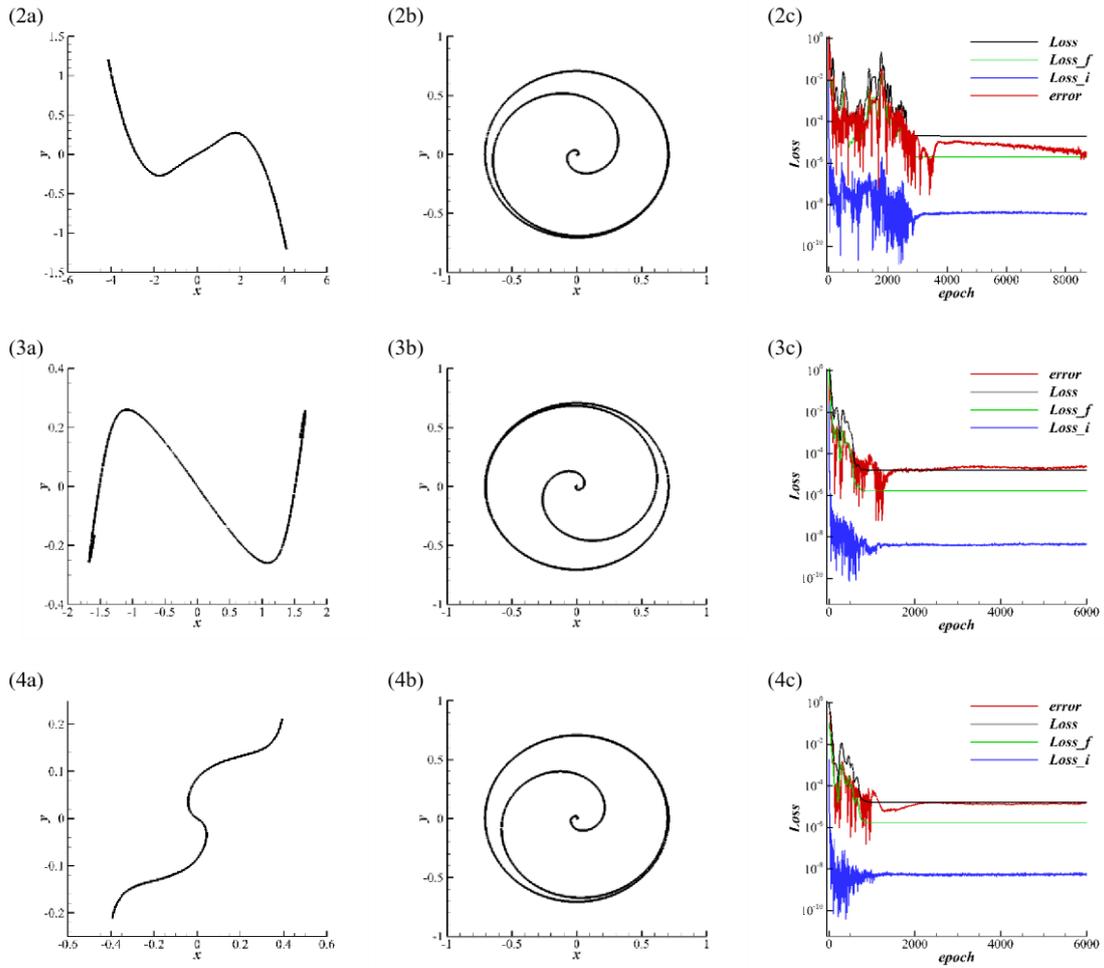

Fig. A5. Four random results of solving the Hopf bifurcation problem using PINNs. The neural network weights are initialized with $R_m = 0.5$. (a) Initial trajectories, (b) final trajectories, and (c) loss and error history.

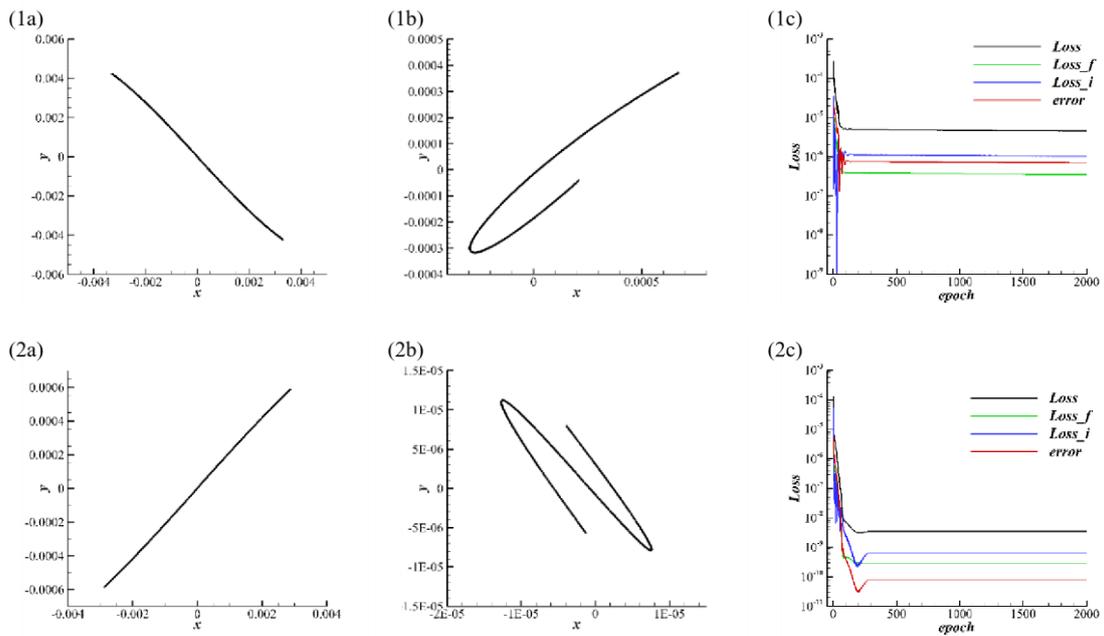

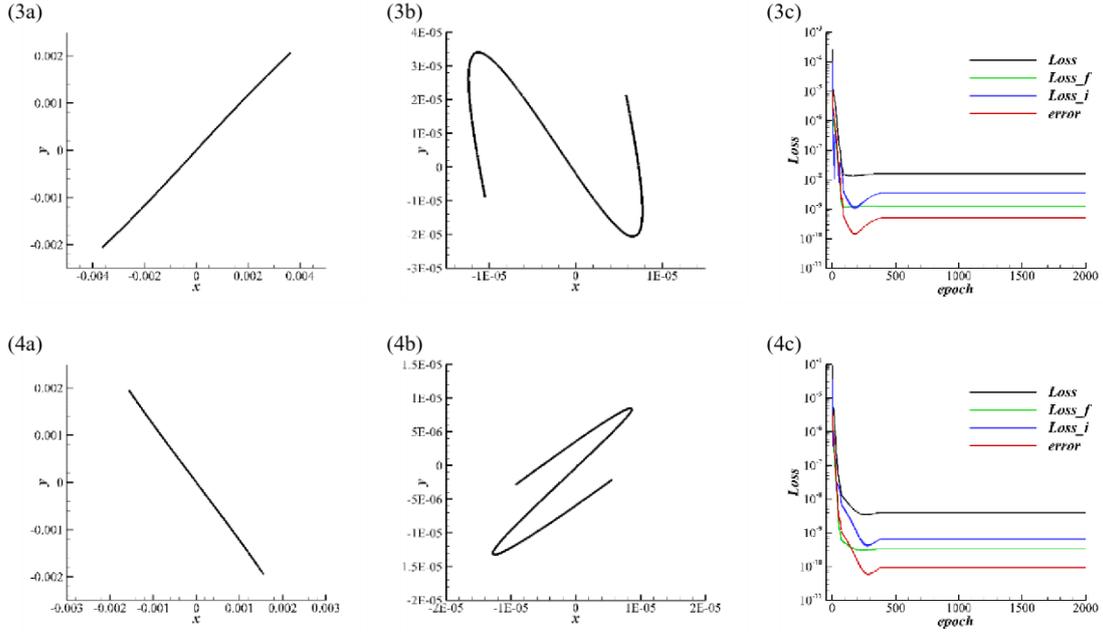

Fig. A6. Four random results of solving the Hopf bifurcation problem using PINNs. The neural network weights are initialized with $R_m = 0.1$. (a) Initial trajectories, (b) final trajectories, and (c) loss and error history.

The numerical results indicate that the optimization process of PINNs exhibits a strong dependence on the initialization scale of the neural network. When the initialization scale is set to $R_m = 0.1$, the initial network outputs have small amplitudes, and the system state remains confined near the origin throughout the training process. Although the loss function continues to decrease, the optimization eventually stagnates at a local stationary point near the origin, failing to approach the limit cycle solution. In contrast, when the initialization range is increased to $R_m = 0.5$, the network outputs exhibit larger initial perturbations, allowing the system state to move away from the origin during training and gradually form a stable periodic orbit. The final solution clearly exhibits a closed limit cycle in phase space, with a radius in good agreement with the theoretical value.

These results demonstrate that, within the automatic differentiation framework of PINNs, the optimization process may converge to different types of solution structures depending on the initial conditions, including unstable equilibrium solutions and stable limit cycles. This observation indicates that, in a residual-minimization-based optimization framework, the convergence behavior is not directly constrained by the stability properties of the original dynamical system. In other words, as long as a state satisfies—or sufficiently reduces—the residual of the governing equations, it may be identified and maintained as a stationary point or local minimum of the optimization problem, regardless of whether it is an attracting state of the original system.

It should be noted that the theoretical analysis of the numerical evolution mechanisms presented in Chapter 4 is primarily based on the Jacobian structure of numerical differentiation PINNs (PINN-ND). In the automatic differentiation PINNs (PINN-AD) framework, however, the explicit Jacobian of the residual with respect to the network parameters is not readily available, making an identical linear stability analysis difficult. Nevertheless, the present Hopf bifurcation example demonstrates that, within the PINN-AD framework, different types of solution structures—including unstable equilibria and stable limit cycles—can still be obtained.

This result indicates that, even under automatic differentiation, the training process of PINNs may converge to qualitatively different dynamical states. From the perspective of a simplified dynamical system, this example confirms that PINN-AD is capable of identifying and maintaining multiple solution branches, thereby providing model-level support for the numerical phenomena observed in the forced oscillating cylinder problem discussed in this work.

## Acknowledgments

We would like to acknowledge the support of the National Natural Science Foundation of China (Grant No.92152301) and the Science and Technology Plan Project of Shaanxi Province (Grant No.2024JC-YBQN-0010).